\documentclass[structabstract]{aa}
\usepackage[draft]{hyperref}
\usepackage{graphicx}
\usepackage{amsmath}
\usepackage{amssymb}
\usepackage{txfonts}               
\usepackage{longtable,lscape}
\usepackage{hyperref}
\usepackage{natbib}                
\bibpunct{(}{)}{;}{a}{}{,}

\newcommand{\hess}{HESS}
\newcommand{\be}{\begin{eqnarray}}
\newcommand{\ee}{\end{eqnarray}}
\begin{document}
\title{A lepto-hadronic model for high-energy emission from FR I radiogalaxies}

\titlerunning{Lepto-hadronic model for FR I radiogalaxies}

\author{
  M.~M.~Reynoso\inst{1}
  \and
  M.~C.~Medina\inst{2}
  \and
  G.~E.~Romero\inst{3}
}
\authorrunning{M.~M.~Reynoso, M.~C.~Medina, \& G.~E.~Romero}
\offprints{M.~C.~Medina \email{\href{mailto:{clementina.medina@cea.fr}}{clementina.medina@cea.fr}}}

\institute{Instituto de Investigaciones F\'{\i}sicas de Mar del Plata (CONICET - UNMdP), Facultad de Ciencias Exactas y Naturales, Universidad Nacional de Mar del Plata, Dean Funes 3350, (7600) Mar del Plata, Argentina \and Irfu, Service de Physique des Particules, CEA Saclay,F-91191 Gif-sur-Yvette Cedex, France \and Instituto Argentino de Radioastronom\'ia, CCT La Plata-CONICET, 1894, Villa Elisa C.C. No. 5, Argentina}

\date{Received 17 May 2010 / 19 April 2011}


    \abstract
    {The well known radiogalaxy Cen\,A has been recently detected as a source of very high energy (VHE) $\gamma$-rays by the \hess\ experiment just before {\it Fermi}/LAT detected it at high energies (HE). The detection, together with that of M87, established radiogalaxies as VHE $\gamma$-ray emitters.}
    {The aim of this work is to present a lepto-hadronic model for the VHE emission from the relativistic jets in  FR I radiogalaxies.}
    {We consider that protons and electrons are accelerated in a compact region near the base of the jet, and they cool emitting multi wavelength radiation as propagating along the jet. The proton and electron distributions are obtained through steady-state transport equation taking into account acceleration, radiative and non-radiative cooling processes, as well as particle transport by convection.}
    {Considering the effects of photon absorption at different wavelengths, we calculate the radiation emitted by the primary protons and electrons, as well as the contribution of secondaries particles ($e^\pm$, $\pi$s and $\mu$s ). The expected high-energy neutrino signal is also obtained and the possibility of detections with KM3NeT and IceCube is discussed.}
    {The spectral energy distribution obtained in our model with an appropriate set of parameters for an extended emission zone can account for much of the observed spectra for both AGNs.}


    \maketitle
    %

    \section{Introduction}

    The non-thermal high-energy emission from Active Galactic Nuclei (AGNs) has been widely studied in recent years at different wavelength ranges through both satellite-borne and ground-based detectors \citep[e.g.][]{1995PASP..107..803U}.
    Several theoretical models have been proposed to explain the electromagnetic emission of these objects. It is commonly accepted that the high-energy radiation is emitted by particles accelerated in the relativistic jets launched from the inner parts of an accretion disk that surrounds the central black hole.
 In general, the high-energy spectral energy distribution (SED) of AGNs presents two characteristic bumps. The lower energy bump, located at optical to X-ray energies, is usually explained as synchrotron emission of electrons while the origin of the high-energy peak in the SED is still under debate {(see e.g. \citep[e.g.][]{2007ASPC..373..169B} for a review)}. Leptonic models attribute this component to inverse-Compton up-scattering off synchrotron or external photons from the disk and/or radiation reprocessed in nearby clouds {\citep[see e.g.][]{2001A+A...367..809K,2008A+A...478..111L,2005A+A...432..401G}}. In hadronic models, interactions of highly relativistic protons in the jet with ambient matter and photon fields, proton-induced cascades, or synchrotron radiation of protons, are responsible for the high-energy photons \citep[see e.g.][]{2000CoPhC.124..290M,2001APh....15..121M,2004A+A...419...89R,2008AIPC.1085..644C,2009MNRAS.393.1041H,1996APh.....5..279R,2009AIPC.1123..242O}. There also exist models which are not based on the emission of accelerated particles in the relativistic jet and  assume the production of TeV $\gamma$-rays in a pulsar-like cascade mechanism in the magnetosphere of the black hole \citep[e.g.][]{2007ApJ...671...85N,2008A+A...479L...5R}.

The recently reported detection by HESS of the nearby radiogalaxy Cen\,A  \citep{2009ApJ...695L..40A} is of great relevance since it establishes radiogalaxies as VHE $\gamma$-ray emitters. Cen\,A is
the second non-blazar AGN discovered at VHE, after the HEGRA detection of $\gamma$-rays from M\,87 \citep{2003A+A...403L...1A} and the later confirmation by HESS \citep{2006Sci...314.1424A}.
A great variety of leptonic and hadronic models has been already applied to this kind of sources and {a full review is} beyond the scope of this work. During the first year of operation, {\it Fermi}/LAT has detected HE emission from Centaurus\,A \citep{2009ApJ...707.1310A} and M\,87 \citep{2009ApJ...707...55A}, providing new constraints to the models.

In this work we present a lepto-hadronic model for the emission from FR I radiogalaxies. Section~\ref{sec:sources} contains a brief description of observational facts on this type of sources. In Section~\ref{sec:scena} we present the outline of our scenario, describing its most relevant characteristics.
Section~\ref{sec:model} is devoted to the description of the model. In Section~\ref{sec:AplCenA} we present the application to Centaurus A, whereas in Section~\ref{sec:AplM87} the results for M87 are given. Finally, in Section ~\ref{sec:concl} we discuss the model implications and perspectives.

\section{FR I radiogalaxies}\label{sec:sources}

According to the unification model of AGNs \citep[e.g.][]{1995PASP..107..803U} FR I radiogalaxies, with their jet axis at a large angle with the line-of-sight, are the parent population of BL Lac objects whose jets are closely aligned to the line of sight. We concentrate here on the only two of them observed until now in the VHE range.

\subsection{Cen\,A} \label{subsec:cenA}
Cen\,A is the closest FR\,I radiogalaxy \citep[$<4$\,Mpc,][]{1984ApJ...287..175H,1993ApJ...414..463H} and its proximity makes it uniquely observable among such objects, eventhough its bolometric luminosity is not high as compared to
other AGNs. It is very active at radio wavelengths presenting a rich jet structure. We can distinguish in its structure two components: inner jets at a kpc scale and giants lobes covering 10$^\circ$ in
the sky. A detailed description of the radio morphology can be found in \cite{1989AJ.....98...27M}. The inner kpc jet has also been detected in X-rays \citep{2002ApJ...569...54K} with an structure of knots and diffuse emission. Recently \cite{2009MNRAS.395.1999C} reported the detection of non-thermal X-ray emission from the shock of the southwest inner radio lobe from deep {\it Chandra} observations.

The supermassive black hole at the center of the active galaxy has an estimated mass of about $10^7 M_\odot$ \citep{2007ApJ...671.1329N,2009MNRAS.tmp..175C,1998A+ARv...8..237I} to $10^8 M_\odot$ \citep{2005AJ....130..406S,2001ApJ...549..915M}. The black hole host galaxy is an elliptical one (NGC\,5128) with a twisted disk which obscures the central engine at optical wavelengths.

Cen\,A was observed by the {\it Compton Gamma Ray Observatory} ({\it CGRO})  with all its instruments from MeV to GeV energies \citep{1992AIPC..254..348G,1995ApJ...449..105K,1993AIPC..280..473P,1998A+A...330...97S,1995ApJS..101..259T}.
In this period, this source exhibited X-ray variability \citep{1996A+A...307..708B} and also some soft $\gamma$-ray variability \citep{1996A+A...307..708B,1995ApJ...449..105K,1998A+A...330...97S}.
However, \cite{1999APh....11..221S} found that the EGRET flux was stable during the whole period of {\it CGRO} observations.

In 1999 the new {\it Chandra} X-ray Observatory took images of Cen\,A with an unprecedented resolution. More than 200 X-ray point sources were identified in those images \citep{2001ApJ...560..675K}.

Cen\,A as a possible source of UHE cosmic rays was early proposed by \citep{1996APh.....5..279R}. Recently, the Pierre Auger Collaboration reported the existence of anisotropy on the arrival directions of UHE cosmic rays \citep{2007Sci...318..938T}, remarking that at least 2 of this events can be correlated with the Cen\,A position ($3^{\circ}$ circle). Further works have claimed that there are several events that can be associated with Cen\,A and its big radio lobes \citep{2008JETPL..87..461G,2008PhyS...78d5901F,2007arXiv0712.3403W} but this correlation is still statistically weak.

Finally, {\it Fermi}/LAT has detected Cen\,A in the first three months of survey with a  significance above 10$\sigma$ \citep{2009ApJ...700..597A}.

    \subsection{M87}\label{subsec:m87}

The giant radiogalaxy M87 is located at 16.7\,Mpc within the Virgo cluster \citep{1999ApJ...521..155M}. It presents a one-sided jet which is inclined with respect to the line of sight an angle between 20$^\circ$ - 40$^\circ$
\citep{1995ApJ...447..582B,1999ApJ...520..621B}. In addition to its bright and well resolved jet, M87 harbors a very massive black hole (6.0 $\pm$ 0.5) $\times$ 10$^9$ M$_\odot$ \citep{2009ApJ...700.1690G}
which is thought to power the relativistic outflow. Given its proximity, the substructures inside the jet could be resolved in the X-ray, optical, and radio wavebands \citep{2002ApJ...568..133W}.
High frequency VLBI observations have resolved the inner jet up to about 70 Schwarzschild radii \citep{1999Natur.401..891J}. Along the jet, nearly stationary components \citep{2008Natur.452..966M} and features moving at superluminal speeds  \citep{2007ApJ...660..200L,2007ApJ...668L..27K} were observed (100 pc-scale).

M87 is also a well-known VHE $\gamma$-rays emitter \citep{2003A+A...403L...1A,2006Sci...314.1424A,2008ApJ...679..397A,2008ApJ...685L..23A} showing a $\gamma$-ray flux variability on short time scales with flaring phenomena in VHE, radio, and X-ray wavebands simultaneously \citep{2009Sci...325..444A}. Recently, it  was detected by {\it Fermi}/LAT with a significance greater than 10$\sigma$ in 10 months of observations \citep{2009ApJ...707...55A}.

Rapid variability constrains the emission region extent to less than $\approx 5 \delta R_{\rm S}$, where $\delta$ is the relativistic Doppler factor. Some suggested explanations for the VHE $\gamma$-ray emission were ruled out (e.g. dark matter annihilation \citep{2000PhRvD..61b3514B}) At the same time, various VHE $\gamma$-ray jet emission models were proposed: leptonic \citep{2005ApJ...634L..33G,2008A+A...478..111L} and hadronic \citep{2004A+A...419...89R} ones. However, the location of the emission region is still unknown. The nucleus \citep{2007ApJ...671...85N,2008IJMPD..17.1569R} , the inner jet \citep{2008MNRAS.385L..98T},  or larger structures in the jet such as the knot HST-1, have been discussed as possible sites of particle acceleration \citep{2007ApJ...663L..65C}.

 \section{Basic Scenario}\label{sec:scena}
We assume that a population of relativistic particles can be accelerated to very high energies close to the base of the AGN jet. These primary electrons and protons carry a fraction of the total kinetic power of the jet $L_{\rm j}^{\rm(kin)}$, and as they are dragged along with the jet, they cool giving rise to electromagnetic emission and neutrinos.

Assuming that a fraction $q_{\rm j}$ of the Eddington luminosity is carried by the jet and a counter jet, the jet kinetic power is
    \be
    L_{\rm j}^{\rm(kin)}= \frac{q_{\rm j}}{2} \frac{4\pi G M_{\rm bh} m_p c}{\sigma_{\rm T}}.
    \ee
This power can be very high if the jet is launched by a dissipationless accretion disk \citep{2010IJMPD..19..339B}.

Most of the jet content is in the form of a thermal plasma with a constant bulk Lorentz factor $\Gamma_{\rm b}$. This plasma is initially in equipartition with a tangled magnetic field at the Alfv$\acute{e}$n surface ($z_0= 50 R_{\rm g}$ from the central black hole) \citep[e.g.][]{2006MNRAS.368.1561M}. The highly disorganized magnetic field has a root mean square value $B(z)$ at a distance $z$ from the black hole in the observer frame, such that $B^2(z):= \langle B_x^2 \rangle+ \langle B_y^2 \rangle+ \langle B_z^2 \rangle$. The magnetic energy density for $z=z_0$ is then
$\rho_{\rm m}(z_0)= B_0^2/(8\pi)$ with $B_0=B(z_0)$. Equating the magnetic to the kinetic energy density, yields:
  \be
    {B_0}= \sqrt{\frac{{8}L_{\rm j}^{\rm(kin)}}{\left[r_{\rm j}(z_0)\right]^2 v_{\rm b} }},
  \ee
where $z_0$ is the distance to the black hole, $v_{\rm b}$ is the jet velocity, and $r_{\rm j}(z_0)=z_0 \tan{\xi_{\rm j}}$ is the radius of the jet assuming that it has a conical shape with half-opening angle $\xi_{\rm j}$. A widely accepted view is that jets are accelerated through the conversion of magnetic energy into kinetic energy \citep[e.g.][]{2007MNRAS.380...51K}. We adopt a phenomenological dependence on the distance to the black hole for the magnetic field \citep[e.g.][]{1999agnc.book.....K},
 \be
   B(z)= B_0\left(\frac{z_0}{z}\right)^m\label{BObs}.
 \ee
Since the density of cold material within the jet decays as $z^{-2}$, using an exponent $m\in(1,2)$ in the above expression implies that, as $z$ increases, the magnetic energy decreases more rapidly than the kinetic one. The corresponding increase in the bulk Lorentz factor is taken into account as described in Appendix \ref{appdx_gam}. In the following, we will write simply $\Gamma_{\rm b}$, but it actually depends on $z$.

The particle acceleration takes place in a compact but inhomogeneous region of size $\Delta z < z_{\rm acc}$ near the base of the jet, at a distance $z_{\rm acc}$  away from the black hole. The value of $z_{\rm acc}$ is fixed by requiring the magnetic energy density to be in sub-partition with the jet kinetic energy density. This condition enables strong shocks to develop \citep{1990cup..book.....G}.
{Assuming that at $z_{\rm acc}$ the magnetic energy is a fraction $q_{\rm m}$ of its value at $z_0$, it follows that}
    \be
    z_{\rm acc}= z_0 q_{\rm m}^{\frac{1}{2-2m}}.
    \ee
For example, if $M_{bh}=10^8 M_\odot$, $m=1.5$, and $z_0=50 R_g$, using $q_{\rm m}=0.38$ yields a distance $z_{\rm acc}=132 R_{\rm g}$.

The power injected in the form of relativistic particles ($L_{\rm rel}$) is considered to be a small fraction $q_{\rm rel}$ of the total jet kinetic power and the relation between proton and electron powers is given by the parameter $a$ such that $L_p= a \ L_e$  \citep[see e. g.][]{2008A+A...485..623R,2010MNRAS.403.1457V}.

These parameters are constrained by the available observational data on each source and reasonable theoretical considerations. In the following section, we describe the procedure used to obtain the particle distributions along the jet and the radiative output that then arises.

The existence of heavy jets magnetically driven in AGN and microquasars has been supported by several scientific works in the last few years. For example, \cite{2010A+A...517A..18S} conclude that mildly relativistic proton-electron jets might be formed by magnetocentrifugal launching by inner portion of magnetized disks around rotating black holes. This leads to a triple-component jet structure: proton-electron component sourrounded by the relativistic pair-dominated sheat. The final speed of this centrifugal outflow depends strongly on the disk vertical structure \citep[see e.g.][]{1993ApJ...410..218W} which is certainly unknown. However, for barionic outflows to reach mildly relativistic speeds, some inital boost would be neccesary, which would be produced by heating or mechanically, by flaring activity and/or by radiation pressure.
Previous work by \cite{2008IJMPD..17.1947H} has proved the existence of a relativistic and baryon-loaded jet in Cygnus X-1 and that the bulk of the kinetic energy is carried by cold protons, as in the case of SS 433.
For further information about magnetically launched jets see \citep[][]{2010LNP...794..233S}, where the production, acceleration, collimation, and composition of jets are well explained.

  \section{Description of the model} \label{sec:model}

The model developed for this work is based on the energy distribution of the different particle populations along the jet. These are obtained as solutions of a 1-dimensional steady-state transport equation that includes the relevant cooling terms and a convective one.
The radiative output is obtained in the jet reference frame, where the particle distributions are isotropic, and the result is transformed back to the observer frame.

The procedure begins with the calculation of the distribution of primary electrons along the jet taking into account synchrotron and adiabatic cooling. After that, the synchrotron radiation emitted by the primary electrons can be calculated. To check that the electron distribution is consistent with the energy loss mechanisms operating, the synchrotron cooling rate must be much greater than the inverse Compton (IC) one due to electrons interacting with the synchrotron photons (SSC). 
If this is the case, it means that the main cooling is due to synchrotron radiation, and neglecting the IC energy loss is a valid approximation to obtain the electron distribution. 
If the SSC cooling can not be neglected, then the transport equation becomes more complicated and a different approach is needed \citep[e.g.][]{2009MNRAS.398.1483S}. 

Having obtained the electron distribution, the next step is to calculate the distribution of primary protons taking into account the cooling due to synchrotron emission, adiabatic expansion, $pp$ and $p\gamma$ interactions. The latter two types of interactions yield the production of secondary pions, muons, and electron-positron pairs. These three populations of particles are also described with the transport equation, and the radiative output that they produce is also considered. According to, e.g. \cite{2008MNRAS.383..467K} and \cite{2010IJMPD..19..671P}, it can be seen that in the present scenario, IC cascades are suppressed by the synchrotron cooling of secondary $e^\pm$, since the magnetic field is greater than $10$ G in the regions of the jet where emission takes place. Therefore, we neglect the effect of IC cascading and calculate the synchrotron emission of the secondary electrons and positrons.

In this section we present all the relevant expressions used in this model. We discuss on the injection of primary particles, the relevant cooling rates, the transport equation used, the injection of secondary particles, and the emission of photons and neutrinos.

  \subsection{Injection of primary particles} \label{subsec:injection}

{At a distance $z$ to the black hole taken in the reference frame of the observer, we adopt an injection function in terms of the particle energy $E'$ in the co-moving reference frame of the jet:}

 \be
   Q'_i(E',z)&:=& \frac{d\mathcal{N}_i} {dE'dV'd\Omega' dt'} \hspace{5mm } ({\rm GeV}^{-1}{\rm s}^{-1}{\rm cm}^{-3}{\rm sr}^{-1})  \label{injdef}\\
&=& K_i\left(\frac{z_{\rm acc}}{z}\right)^2{E'}^{-s}\exp{\left[-\left(\frac{E'}{{E'}^{({\rm max})}_{i}}\right)\right]}.\label{Qep}
 \ee

Here, $\mathcal{N}_i$ represents the total number of relativistic electrons ($i=e$) or protons ($i=p$), and the cut-off energy is obtained from the balance of particle gains and losses. These processes are described below. 

The injection function can be transformed to the observer frame by taking into account that 
\be
 \frac{E'}{{p'}^2} \frac{d\mathcal{N}_i}{dV' dp' d\Omega' dt'} 
\ee
is a Lorentz invariant \citep[e.g.][]{2002ApJ...575..667D,2011A+A...528L...2T}. From this, it follows that the particle injection in the observer frame is given by
 \be
  Q_i(E,z)= \left[\frac{E^2- m_i^2c^4}{{E'}^2- m_i^2c^4}\right]^{1/2} Q_i'(E',z'),\label{Qjet2obs}
 \ee
{where the energy in the jet frame is given in terms of the energy in the observer frame $E$, and the angle $\theta$ between the particle momentum $\vec{p}$ and the bulk velocity of the jet $\vec{v}_{\rm b}= c \beta_{\rm b} \hat{z}$:} 
 \be
   E'= \Gamma_{\rm b}\left(E- \beta_{\rm b} \cos{\theta} \sqrt{E^2-m_i^2c^4}\right).
 \ee
 
The normalization constant $K_i$ is found in each case using the power in relativistic species $L_{\{e,p\}}$, integrating in the energy, solid angle, and volume of the acceleration zone ($z\in[z_{\rm acc},z_{\rm acc}+\Delta z]$):
 \be
 L_i=  \int_{\Delta E} dE\int_{4\pi} d\Omega \int_{\Delta V} dV \; E \; Q_i(E,z).
 \ee

 \subsection{Accelerating and cooling processes}\label{subsec:acc_cool}

To determine the maximum energies, it is necessary to account for the particles acceleration and cooling rates ${t'}^{-1}={E'}^{-1}|dE'/dt'|$.

We assume that in the acceleration zone, particles are accelerated by a diffusive shock-modulated mechanism with a rate \citep[e.g.][]{1990ApJ...362...38B}:
 \be
  {t'}_{\rm acc}^{-1}(E',z) =  \eta \frac{c e  B'(z)}{E'} \label{tacc},
 \ee
where $0<\eta\ll 1$ is the efficiency of the mechanism. The magnetic field in the jet frame is assumed to be random and with no preferred direction. $B'$ represents the root mean square field in the jet frame: $ {{B'}^2= \langle{B'_{x'}}^2\rangle+ \langle{B'_{y'}}^2\rangle + \langle{B'_{z'}}^2\rangle}$. {As considered also in \cite[][]{2000ApJ...535..104H}, the average of each component of the magnetic field is supposed to vanish in the jet frame: ${\langle B'_i\rangle=0}$. }
Assuming that there is no electric field in the jet frame, the components of the magnetic field in the observer frame are related to the ones in the jet frame as
 $B_x= \Gamma_{\rm b} B'_{x'}$, $B_y= \Gamma_{\rm b} B'_{y'}$, and $B_z= B'_{z'}$, which yields\footnote{{Note that Eq.(\ref{Btransf}) is consistent with Eq. (3) of \cite[][]{2000ApJ...535..104H} if the radial component of their flow velocity can be neglected.}}
  \be
  B'(z)= \sqrt{\frac{3}{2\Gamma_{\rm b}^2 + 1}} B(z).\label{Btransf}
  \ee
{where $B(z)$ is given by Eq.(\ref{BObs}).}

As for the energy loss processes, particles emit synchrotron radiation at a rate
 \be
{t'}_{\rm sync}^{-1}(E',z)=\frac{4}{3}\left(\frac{m_e}{m}\right)^3\frac{
\sigma_{\rm T}{B'}^2}{m_e c \ 8\pi}\frac{E'}{m c^2}\label{tsyn}.
 \ee
The lateral expansion of the jet implies an adiabatic cooling rate \citep{2006A+A...447..263B} in the observer frame given by
 \be
 {t}_{\rm ad}^{-1}(E,z)= \frac{2}{3}\frac{v_{\rm b}}{z}.
 \ee

The density of cold matter at a distance $z$ from the black hole is, in observer frame,
 \be
n_{\rm c}(z)= \frac{(1-q_{\rm rel})\dot{m}_{\rm j}}{m_p  \pi z^2\tan^2{\xi_{\rm j}} v_{\rm b} },
 \ee
where the mass loss rate in the jet is $$\dot{m}_{\rm j}=\left.\frac{L_{\rm j}^{\rm (kin)}(\Gamma_{\rm b}-1)}{c^2}\right|_{z_0}.$$
In the jet frame, the cold matter density is $$n'_{\rm c}(z)=\frac{n_{\rm c}(z)}{\Gamma_{\rm b}}.$$	

Relativistic protons in the jet undergo $pp$ collisions with these cold protons at a rate
 \be
{t'}_{pp}^{-1}(E',z)= {n'_{\rm c}(z)} \; c \; \sigma_{pp}^{\rm(inel)}(E')K_{pp}.
 \ee

Here the inelasticity coefficient is $K_{pp}\approx 1/2$, the corresponding cross section for inelastic $pp$ interactions can be approximated by \citep{2006PhRvD..74c4018K}\citep{2009PhRvD..79c9901K}

\begin{multline}
\sigma_{pp}^{\rm(inel)}(E_p)= (34.3+ 1.88 L+ 0.25 L^2) \\ \times
\left[1-\left(\frac{E_{\rm th}}{E_p}\right)^4 \right]^2 \times
10^{-27}{\rm cm}^2,
\end{multline}

where $L= \ln(E_p/1000{\rm \ GeV})$ and $E_{\rm th}=1.2 {\rm \
GeV}$.

\subsubsection{Synchrotron radiation and Inverse Compton interactions}\label{subsubsec:IC_int}

To consider IC interactions of primary electrons with synchrotron photons in the jet (SSC), it is necessary to know the particle distribution of the synchrotron emitting particles, $N'_{e,p}(E',z)$, which is obtained by solving the transport equation described below. Actually, the synchrotron emission of electrons dominates the low energy photon background which is a good target for IC and $p\gamma$ interactions. Since in the cases considered here, IC cooling is much less efficient than synchrotron cooling, the electron distribution can be obtained to a good approximation without considering IC cooling. In the jet frame, the background radiation density, in units [${\rm GeV}^{-1}{\rm cm}^{-3}$], has been approximated locally as
 \be
 n'_{\rm ph}(E'_{\rm ph},z)\approx \frac{\varepsilon'_{\rm syn}(E'_{\rm ph},z)}{E'_{\rm ph}}
\frac{r_{\rm j}(z)}{c},
 \ee
where $\varepsilon'_{\rm syn}$ is the power per unit energy per unit volume of the synchrotron photons,
 \be
   \varepsilon'_{\rm syn}(E'_{\rm ph},z)=
   {\left(\frac{1- e^{-\tau_{\rm SSA}(E'_{\rm ph},z)}}{\tau_{\rm SSA}(E'_{\rm ph},z)} \right)}\int_{m_e c^2}^{\infty} dE' 4\pi  P_{\rm syn}
    N'_{e}(E',z).\label{varepsilon}
 \ee
Here, the synchrotron power per unit energy emitted by the electrons is given by \citep{1970RvMP...42..237B}:
  \be
 P_{\rm syn}(E'_{\rm ph},E',z)= \frac{\sqrt{2} e^3 B'(z)}{m_e c^2 h}
\frac{E'_{\rm ph}}{E_{\rm cr}} \int_{E'_{\rm ph}/E_{\rm cr}}^\infty d\zeta
K_{5/3}(\zeta), \ee
 where $K_{5/3}(\zeta)$ is the modified Bessel function of order $5/3$ and
$$E_{\rm cr}= \frac{\sqrt{6}he B'(z)}{4\pi m_e c}\left(\frac{E'}{m_e
c^2}\right)^2.$$ 

{The factor in parenthesis in Eq. (\ref{varepsilon}) accounts for the effect of synchrotron self-absorption (SSA) within the jet, with an optical depth 
$$\tau_{\rm SSA}(E'_{\rm ph},z)= \int_0^{\frac{r_{\rm j}(z)}{\sin{\theta'}}} dl' \alpha_{\rm SSA}(E'_{\rm ph},z'(l')), $$ where $\theta'$ is such that $\tan{\theta'}= \Gamma_{\rm b}\tan{\theta}$, and the SSA coefficient is given by
\citep{1979rpa..book.....R}:} 
\begin{multline} \alpha_{\rm SSA}=
-\frac{h^3c^2}{8\pi {E'_{\rm ph}}^2}\int_{m_e c^2}^\infty dE' {E'}^2 P_{\rm
syn}(E'_{\rm ph},E')\\ \times
  \frac{\partial}{\partial
E'}\left[\frac{N'_e(E',z'(l'))}{{E'}^2}\right].\end{multline}

After all these considerations, the cooling rate due to IC scattering for electrons of energy $E'$ {in the jet frame} can be obtained by integrating in the {target} photon energy $E'_{\rm ph}$ and in energy of the scattered photon $E'_\gamma$ \citep{1970RvMP...42..237B}:
 \begin{multline}
 {t'}^{-1}_{\rm IC}(E',z)= \frac{3 m_e^2 c^4 \sigma_{\rm T}}{4 {E'}^3}\int_{{E'}_{\rm ph}^{\rm(min)}}^{E'}dE'_{\rm ph } \frac{n'_{\rm ph}(E'_{\rm ph},z)}{{E'}_{\rm ph}} \\ \times \int_{{E'}_{\rm ph}}^{\frac{\Gamma_e}{\Gamma_e+1}E'} dE'_\gamma  F(q)\left[E'_\gamma-E'_{  \rm ph}\right], \label{tIC}
 \end{multline}
where {${E'}_{\rm ph}^{\rm(min)}$ is the lowest energy of the available background of synchrotron photons, and}
 \be
 F(q)= 2q \ln q + (1 + 2q) (1 - q)+ \frac{1}{2}(1 - q)\frac{(q \Gamma'_{\rm e})^2}{1 + \Gamma'_{\rm e} },
 \ee
with $\Gamma'_{\rm e}=4 E'_{\rm ph}E'/(m_e^2c^4)$ and
 $$q=\frac{E'_\gamma}{\Gamma'_{\rm e}E'_{\rm ph}(1-{E'_\gamma/}{E'_{\rm ph}})}. $$

\subsubsection{Proton-photon interactions}\label{subsubsec:pg_int}

 Proton interactions with the photon background can be an important cooling process, with a rate appoximated as
  \begin{multline}
{t'}_{p\gamma}^{-1}(E',z)=\frac{c}{2\gamma_p^2}\int_{\frac{\epsilon_{\rm
th}^{(\pi)}}{2\gamma_p}}^{\infty} dE'_{\rm ph}\frac{n'_{\rm
ph}(E'_{\rm ph},z)}{{E'}_{\rm ph}^2}  \\
\times \int_{\epsilon_{\rm th}^{(\pi)}}^{2E'_{\rm ph}\gamma_p} d \epsilon_{\rm r}
  \sigma_{p\gamma}^{(\pi)}(\epsilon_{\rm r})K_{p\gamma}^{(\pi)}(\epsilon_{\rm r})
  \; \epsilon_{\rm r}. \label{tpg}
\end{multline}
Here, $\epsilon_{\rm th}^{(\pi)}=150$ MeV and we use the expressions for the cross section $\sigma_{p\gamma}^{(\pi)}$
and the inelasticity $K_{p\gamma}^{(\pi)}$ given in \cite{1990ApJ...362...38B}.

\subsection{Particle distributions in the jet}\label{subsec:part_dist}
The energy distribution for each particle population is obtained as a solution to the following $1$-D stationary transport equation \citep[e.g.][]{1964ocr..book.....G,2008MNRAS.383..467K}:
 \be
 v_{\rm b}\frac{\partial N'(E',z)}{\partial z}+ \frac{\partial\left[ b(E',z) N'(E',z)\right]}{\partial E'} + \frac{N'(E',z)}{T_{\rm d}(E') }=
 Q'(E',z).\label{transport_eq}
 \ee
This equation includes the effects of convection with a speed $v_{\rm b}\sim c$, and particle cooling with an energy loss $b(E',z)=dE'/dt$. Particle decay with a timescale $T_{\rm d}(E')$ is also considered in the cases of secondary pions and muons. {We note that the energy $E'$ in the above equation corresponds to the co-moving frame, where the particle distributions are isotropic, while the spatial coordinate $z$ corresponds to the observer frame \citep[e.g.][]{1970ApJ...160..735J,1988ApJ...328..269K}}. 
The source term is given by a function $Q'(E',z)$, which in the case of primary particles is given by Eq. (\ref{Qep}), while for secondaries it is obtained using the parent particle distribution ($N'(E',z)$) along with the secondary particle production rates (see Section~\ref{subsubsec:pimu_prod}).

We can solve the transport equation using the method of characteristics, i.e., writing
 \be
\frac{dz}{v_{\rm b}}=\frac{dE'}{b(E',z)}=\frac{dN'(E',z)}{ Q'(E',z)- \left[\frac{1}{T_{\rm d}(E') }+\frac{\partial
b(E',z)}{\partial E}\right] N(E,z) },
 \ee
where the first two terms allow us to find a characteristic curve $z_c(E_c)$ for each pair $(E',z)$ of interest. Equating the second and third members it follows that
\begin{multline}
N'(E',z)= \int_E^{\infty} dE_c\frac{Q'(E_c,z_c)}{|b(E_c,z_c)|}\\ \times \exp{\int_{E'}^{E_c}dE''\left[\frac{T_{\rm d}(E'')|\frac{db}{dE}(E'',z'')|-1}{T_{\rm d}(E'')|b(E'',z'')|}\right]}.
\end{multline}


\subsubsection{Pion and muon production}\label{subsubsec:pimu_prod}

As high energy protons interact with background matter and radiation, they produce pions.

The pion injection due to $pp$ interactions in the jet frame is calculated as
\begin{multline}
Q'_{\pi, pp}(E_\pi',z)=n'_{\rm c}(z)\; c\int_{0}^{1}  \frac{dx}{x} N_p' \left(\frac{E_\pi'}{x},z\right)
\\ \times
F_\pi\left(x,\frac{E_\pi'}{x}\right)\sigma_{pp}^{\rm(inel)} \left(\frac{E_\pi'}{x}\right)
\label{Qpipp}
\end{multline}
where
\begin{multline}
F_\pi\left(x,\frac{E_\pi'}{x}\right)=4\alpha B_\pi
x^{\alpha-1}\left(\frac{1-x^\alpha}{1+r'
x^\alpha(1-x^\alpha)}\right)^4 \\ \times \left(
\frac{1}{1-x^\alpha}+ \frac{r'(1-2x^\alpha)}{1+
r'x^\alpha(1-x^\alpha)} \right) \left(1-\frac{m_\pi c^2}{
E_\pi'}\right)^{1/2}
\end{multline}
is the distribution of pions produced per $pp$ collision, with $x=
E'_\pi/E'$, $ B_\pi=a'+ 0.25$, $a'= 3.67+ 0.83 L+ 0.075 L^2$, $r'=
2.6/\sqrt{a'}$, and $\alpha= 0.98/\sqrt{a'}$  \citep[see][]{2006PhRvD..74c4018K,2009PhRvD..79c9901K}.

In the same way, the source function for charged pions produced by $p\gamma$ interactions is
\begin{multline}
Q'_{\pi, p\gamma}(E_\pi',z)= \int_{E_\pi'} dE' N'_p(E',z) \;
{\omega'}_{p\gamma}^{(\pi)}(E',z) \\ \times
\mathcal{N}_\pi(E') \; \delta(E'_\pi-0.2E') \\
= 5 \; N'_p(5E_\pi',z) \; {\omega'}^{(\pi)}_{p\gamma}(5E_\pi',z) \;
\mathcal{N}_\pi(5E_\pi').\label{Qpipg}
\end{multline}
Here ${\omega'}_{p\gamma}^{(\pi)}$ is the $p\gamma$ collision frequency defined as \citep{2003ApJ...586...79A}:
\begin{multline}
 {\omega'}_{p\gamma}^{(\pi)}(E',z)= \frac{c}{2\gamma_p^2} \int_{\frac{\epsilon_{\rm
th}^{(\pi)}}{2\gamma_p}}^{\infty} dE_{\rm ph}'\frac{n'_{\rm
ph}(E_{\rm ph}',z)}{{E_{\rm ph}'}^2}
\\  \times \int_{\epsilon_{\rm
th}^{(\pi)}}^{2\epsilon\gamma_p} d \epsilon'
  \sigma_{p\gamma}^{(\pi)}(\epsilon')
  \; \epsilon',
 \end{multline}
and the mean number of positive or negative pions is
 \be
\mathcal{N}_\pi \approx \frac{p_1}{2} + 2p_2.
 \ee
This number depends on the probabilities of single pion and multi-pion production $p_1$ and $p_2=1-p_2$. Using the mean
inelasticity function $\bar{K}_{p\gamma}={t_{p\gamma}^{-1}/\omega_{p\gamma}^{(\pi)}}$, the probability $p_1$ is
\be
p_1=\frac{K_2-\bar{K}_{p\gamma}}{K_2-K_1}, \ee where $K_1=0.2$ and
$K_2=0.6$.

The injection functions of charged pions given by Eqs. (\ref{Qpipp},\ref{Qpipg}) {are used to work out}
the distribution $N'_\pi(E',z)$ by solving the transport equation (\ref{transport_eq}).

For muon injection, we proceed as \citep[][]{2007PhRvD..75l3005L} and we consider the production of left handed and right handed muons separately, which have different decay spectra:
\be \frac{dn_{\pi^- \rightarrow
\mu^-_L}}{dE_\mu}(E_\mu;E_\pi)= \frac{r_\pi(1-x)}{E_\pi
x(1-r_\pi)^2}\Theta(x-r_\pi) \\
\frac{dn_{\pi^- \rightarrow \mu^-_R}}{dE_\mu}(E_\mu;E_\pi)=
\frac{(x-r_\pi)}{E_\pi x(1-r_\pi)^2}\Theta(x-r_\pi),
\ee
with $x= E_\mu/E_\pi$ and $r_\pi= (m_\mu/m_\pi)^2$.

The injection function of negative left handed and positive right handed muons is
\begin{multline}
  Q'_{\mu^-_L,\mu^+_R}(E'_\mu,z)= \int_{E'_\mu}^{\infty} dE'_\pi
T_{\pi,{\rm d}}^{-1}(E'_\pi)\\
\times\left(  N'_{\pi^-}(E'_\pi,z) \frac{dn_{\pi^- \rightarrow\mu^-_L}}{dE'_\mu}(E'_\mu;E'_\pi) \right.\\ \left.
  + N'_{\pi^+}(E'_\pi,z)\frac{dn_{\pi^+ \rightarrow
  \mu^-_R}}{dE'_\mu}(E'_\mu;E'_\pi) \right).
\end{multline}
Given that CP invariance implies that $dn_{\pi^- \rightarrow\mu^-_L}/dE'_\mu= dn_{\pi^+ \rightarrow \mu^+_R}/dE'_\mu$, and since
the total distribution obtained for all charged pions is
$N'_\pi(E'_\pi,z)= N'_{\pi^+}(E'_\pi,z)+ N'_{\pi^-}(E'_\pi,z)$, it follows
that the injection for left handed muons is
\begin{multline}
  Q'_{\mu^-_L,\mu^+_R}(E'_\mu,z)= \int_{E'_\mu}^{\infty} dE'_\pi T_{\pi,
{\rm d}}^{-1}(E'_\pi)
  \\ \times
  \ N'_{\pi}(E'_\pi,z) \frac{dn_{\pi^- \rightarrow
  \mu^-_L}}{dE'_\mu}(E'_\mu;E'_\pi). \label{QmuL}
\end{multline}
And in a similar way, the right handed muons injection is
\begin{multline}
  Q'_{\mu^-_R,\mu^+_L}(E'_\mu,z)= \int_{E'_\mu}^{\infty} dE'_\pi T_{\pi,
{\rm d}}^{-1}(E'_\pi)
  \\ \times
  \ N'_{\pi}(E'_\pi,z) \frac{dn_{\pi^- \rightarrow
  \mu^-_R}}{dE'_\mu}(E'_\mu;E'_\pi). \label{QmuR}
\end{multline}

Again, the injection functions of muons given by Eqs. (\ref{QmuL},\ref{QmuR}) {are used to work out the}
distribution $N'_\mu(E',z)$ by solving the transport equation (\ref{transport_eq}).

\subsection{Electromagnetic radiation}\label{subsec:elect_emis}

The main radiative processes considered in this work are: synchrotron radiation, IC emission, $pp$ and $p\gamma$ interactions.
For each emission process we can calculate the injection of photons or radiation emissivity, which is given by the general expression of Eq. (\ref{injdef}) and represents the number of photons produced per time unit, per volume unit, per photon energy, and per solid angle.
{In the observer frame, the corresponding emissivity is given by}
 \be
 Q_\gamma(E_\gamma,z)= D \, {Q'_\gamma(E'_\gamma,z)},\label{Qgtrans}
 \ee
 {where the Doppler factor is $D=\Gamma_{\rm b}^{-1}(1-\beta_{\rm b}\cos\theta)^{-1}$ and $ E_\gamma= D\ E'_\gamma $.}

In the case of synchrotron radiation, the emissivity in the jet frame is
\be
Q'_{\gamma,{\rm syn}}(E'_\gamma,z)= \frac{\epsilon'_{\rm syn}(E'_{\rm ph},z)}{4\pi E'_\gamma}.
\ee

and the IC emissivity is

 \begin{multline}
Q'_{\gamma, {\rm IC}}(E'_\gamma,z)= \frac{r_e^2 c}{2}\int_{{E'}_{\rm ph}^{\rm(min)}}^{E'_\gamma}dE'_{\rm ph} \frac{n'_{\rm ph}(E'_{\rm ph},z)}{E_{\rm ph}} \\
  \times \int_{E'_{\rm min}}^{E'_{\rm max}}d{E'} \frac{N'_e({E'},z)}{\gamma_e^2}F(q),
  \end{multline}
where {$r_e$ is the classical electron radius, and} we integrate in the target photon energy $E'_{\rm ph}$ 
and in the electron energies $E'$ between
 \be
 E'_{\rm min}= \frac{E'_\gamma}{2}+\frac{m_e c^2}{2}\sqrt{\frac{E'_\gamma}{2E'_{\rm ph}}+\frac{{E'}_\gamma^2}{2m_e^2c^4}}
  \ee
 and
 \be
 E'_{\rm max}= \frac{E'_\gamma}{1-\frac{E'_\gamma}{E'_{\rm ph}}}.
 \ee

With respect to the hadronic contribution to the total emission, the photon emissivity due to $pp$ interactions is obtained as

 \begin{multline}
Q'_{\gamma, pp}(E'_\gamma,z)= n'_{\rm c}(z) c \int_{0}^{1}\frac{dx}{x}N'_p\left(\frac{E'_\gamma}{x},z\right)
 \\ \times
F_\gamma\left(x,\frac{E'_\gamma}{x}\right)\sigma_{pp}^{\rm(inel)}\left(\frac{E'_\gamma}{x}\right),
  \end{multline}
where the function $F_\gamma(x,E')$ is the same as defined  by \cite{2006PhRvD..74c4018K}, for a proton energy $E'=E'_\gamma/x$.

For the $\gamma$-ray emissivity due to $p\gamma$ interactions we use the expression
\begin{multline}
Q'_{\gamma, p\gamma}(E'_\gamma,z)= \frac{1}{4\pi}\int_{E'_\gamma}\frac{dE'}{E'} N'_p(E',z)\\
  \times\int_{E_{\rm ph}^{\rm (min)}}^{E'_\gamma}dE_{\rm ph}' n'_{\rm ph}(E_{\rm ph}',z)\Phi\left(\eta_{p\gamma},x\right),
\end{multline}
where $E'_\gamma=x E'$, $\eta_{p\gamma}=\frac{4E'{E}_{\rm ph}'}{m_p^2c^4x}$,
\be
E_{\rm ph}^{\rm (min)}= \frac{m_p^2c^4x}{4E'_\gamma}\left[2\left(\frac{m_{\pi_0}}{m_p}\right) +\left(\frac{m_{\pi_0}}{m_p}\right)^2\right],
\ee

and with the function $\Phi\left(\eta_{p\gamma},x\right)$ as tabulated in \cite{2008PhRvD..78c4013K}.

\subsection{Internal and external photon absorption}\label{subsec:abs}

We consider the absorption of the radiation due to photoionization processes at eV energies, and also due to $e^-e^+$ creation by $\gamma\gamma$ and $\gamma N$ interactions at high energies in the source.

\begin{figure}[!htp]
\includegraphics[trim = 6mm 4mm 4mm 10mm, clip,width=\linewidth,angle=0]{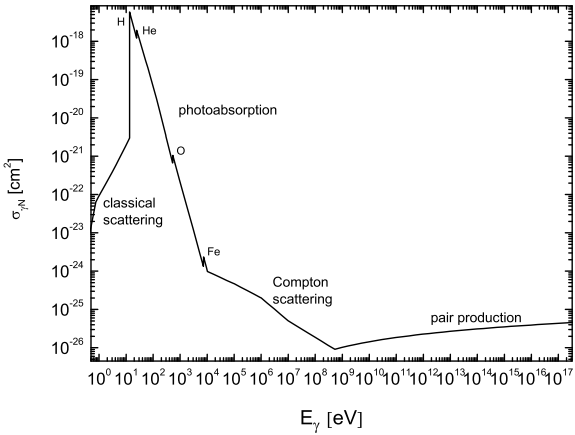}
\caption{Cross section for $\gamma N$ interactions. {The absorption edges corresponding to the different elements are indicated.}}
\label{Fig:sigmaGH}
\end{figure}

 For $\gamma N$ interactions, we take as target the material along the line of sight corresponding to each particular object, which is measured through the column density {of neutral hydrogen} $N_H$. The absorption cross section $\sigma_{\gamma N}$ in this case is taken {as in \cite{1996Ap+SS.236..285R} for $E_\gamma<1$ keV, assuming that the medium is composed by atomic hydrogen and dust within galactic abundances (see Fig. \ref{Fig:sigmaGH}). This cross section includes the effects of photoionizaton for $E_\gamma>13.6$ eV and scattering with dust below this energy. In the present work we do not consider recombination, the inverse process of photoionization (see Appendix \ref{appdx_rec} )}. The resulting optical depth for $\gamma N$ interactions is {approximated as}
 \be
   \tau_{\gamma H}(E_\gamma)= N_H \sigma_{\gamma N}(E_\gamma).
 \ee

{It can be seen from Fig. \ref{Fig:sigmaGH} that besides the large absorption edge corresponding to the ionization energy of hydrogen, three aditional absorption edges are included in the cross section. They correspond to different ionization energies of helium, K-shell electrons of oxygen, and iron. In the present context, as can be seen below, we find no significant features associated with these aditional absorption edges \citep[e.g.][]{1974ApJ...187..497C, 1999ApJ...517..168G}. 
The possible re-emission of lines is beyond the scope of this work, and could be studied, e.g., as in \cite{1999ApJ...517..168G}.}
{For energies above $1$ keV we take the $\gamma N$ cross section from \cite{2008PhLB..667....1P}, to account for Compton scattering and $e^\pm$ production. In Fig. \ref{Fig:sigmaGH} we show the cross section used for $\gamma\,N$ interactions}

In the case of $\gamma \gamma$ interactions, the main radiation target is considered to be the synchrotron photons inside the jet, which are characterized by a radiation density $n'_{\rm ph}(E'_{\rm ph},z)$. If a dissipationless accretion disk is present, no other significant photon field is relevant for gamma-ray absorption.
The cross section is given by
 \begin{multline}
\sigma_{\gamma\gamma}(E'_\gamma,E'_{\rm ph})=\frac{3 \sigma_{\rm T}}{16}(1-\beta_e^2)
\times \nonumber\\ \left[2\beta_e(\beta_e^2-2)+
(3-\beta_e^4)\ln\left(\frac{1+\beta_e}{1-\beta_e}\right)\right],
 \end{multline}
with
 \be
 \beta_e=\left[1-\frac{2m_e^2 c^4}{E'_\gamma E'_{\rm ph}(1-\cos{\theta_{\gamma\gamma}})}\right]^{1/2},
 \ee
{where $\theta_{\gamma\gamma}$ is the angle of interaction between the incident photons.  For $\gamma$-rays produced at a position $z$ in the jet, we assume that they undergo collisions over a length $r_{\rm j}(z)/\sin{\theta'}$ inside the jet with a target radiation field that is isotropic in the jet frame. The corresponding optical depth is calculated using the general expression of  \cite{1967PhRv..155.1404G} to obtain}
 \begin{multline}
 \tau_{\gamma\gamma}^{\rm(int)}(E'_\gamma,z)=2\pi\int_{0}^{\frac{r_{\rm j}(z)}{\sin{\theta'}}} dl' \int_{-1}^1dx'(1-x') \\ \times \int_{\frac{2m_e^2c^4}{E'_\gamma(1-x')}}^{\infty} dE'_{\rm ph} n'_{\rm ph}(E'_{\rm ph},z'(l')) \sigma_{\gamma\gamma}(E'_\gamma,E'_{\rm ph}).
 \end{multline}

\begin{figure}
\includegraphics[trim = 6mm 5mm 4mm 5mm, clip,width=\linewidth,angle=0]{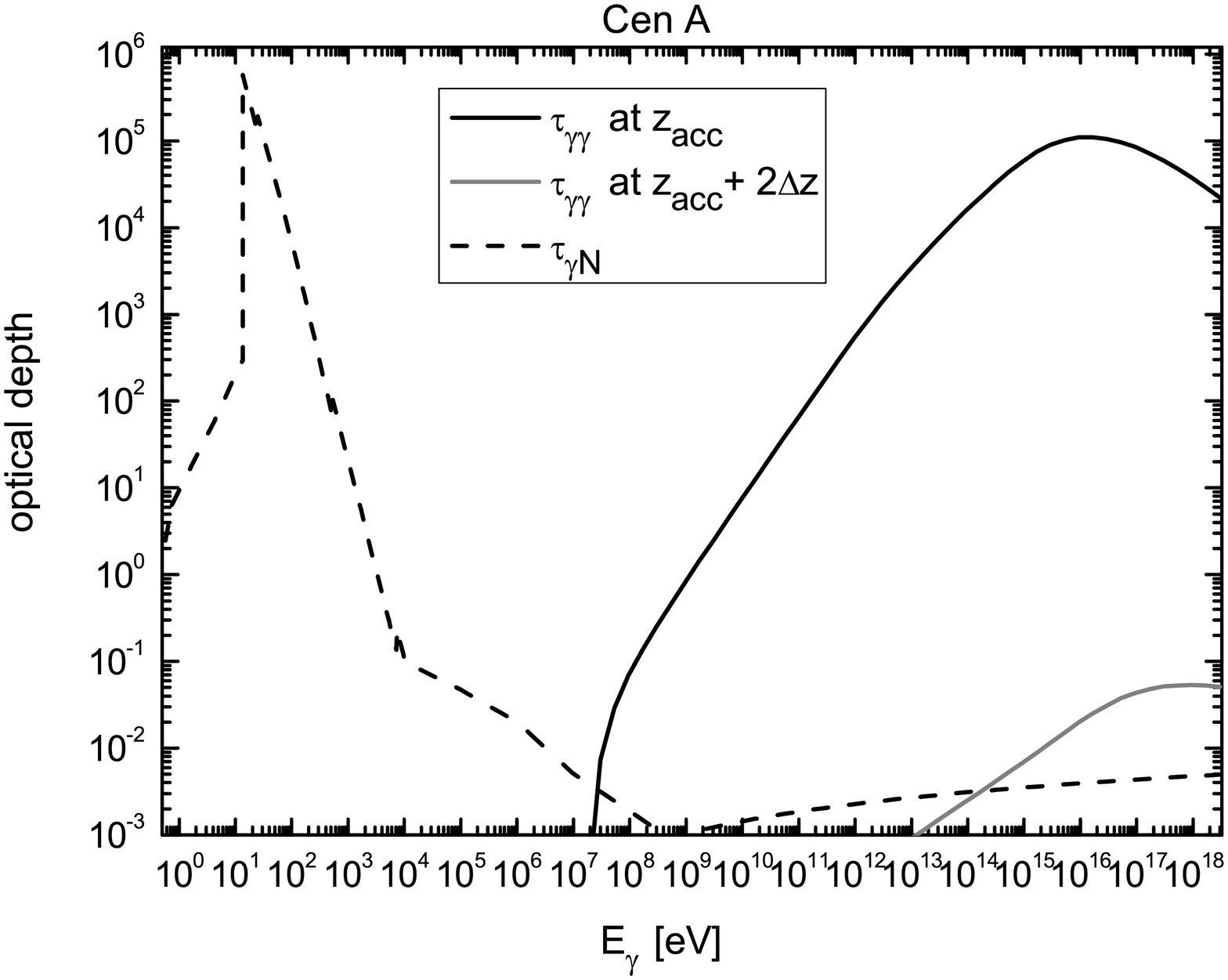}
\includegraphics[trim = 6mm 5mm 4mm 5mm, clip,width=\linewidth,angle=0]{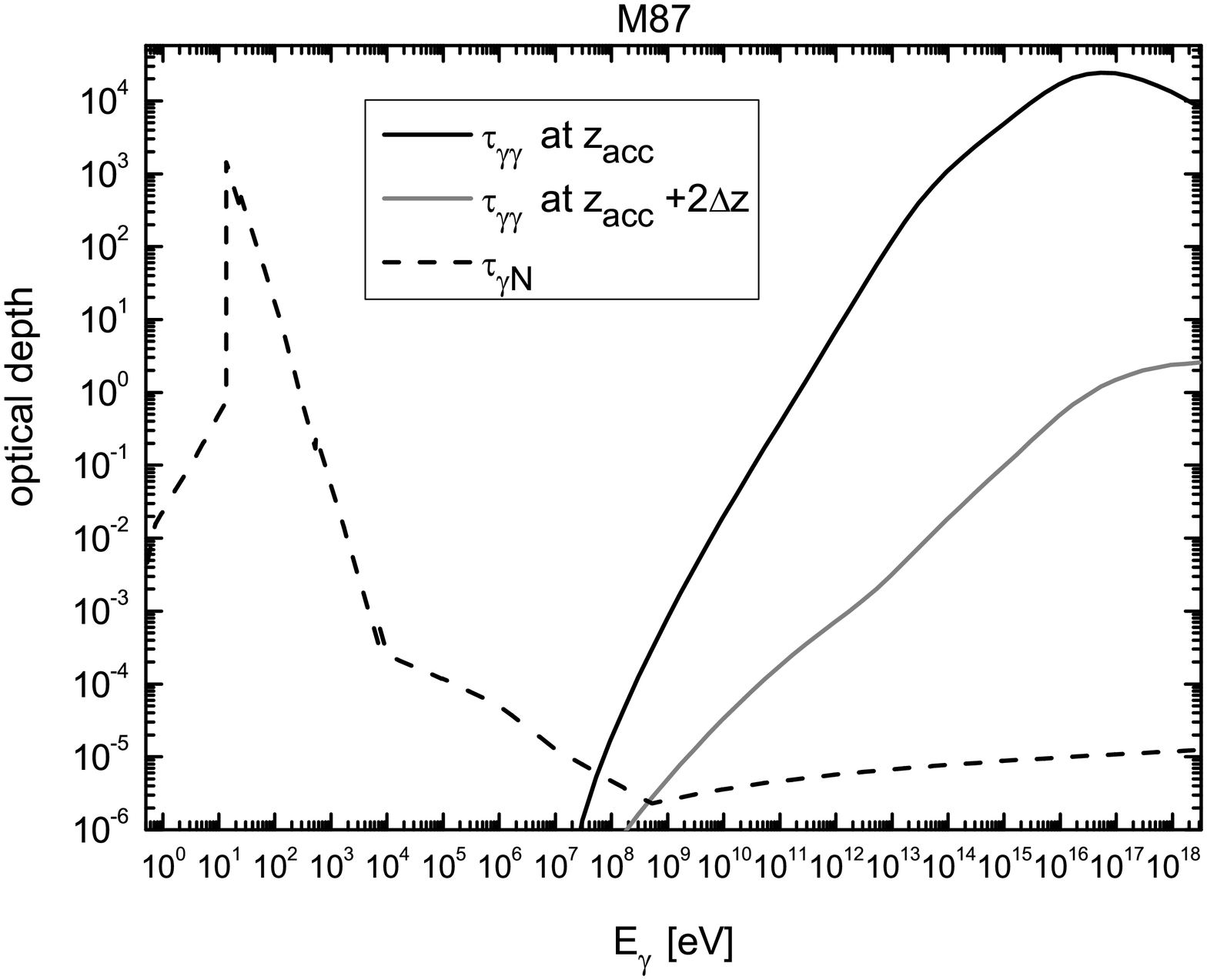}
\caption{Optical depth for Cen A and M87 at different distances form the black hole.}
\label{Fig:tauCena}
\end{figure}

In Fig. \ref{Fig:tauCena} we show the optical depths obtained for the case of Cen A {and M87}, $\tau_{\gamma N}$ and $\tau_{\gamma\gamma}$, the latter evaluated within the injection zone and also outside it. The absorbing photons outside the injection zone are those of synchrotron emission of protons and secondary $e^\pm$, which in that case are also included in $n'_{\rm ph}(E'_{\rm ph},z)$. It can be seen from this plot that very important absorption occurs at low energies by photoionization, and also at VHE through internal $\gamma\gamma$ interactions. The latter effect takes place mainly at the injection zone, since for $z> z_{\rm acc}+ \Delta z$, the optical depth is much lower.

Taking into account the total optical depth $\tau_{\gamma}(E_\gamma,z)=\tau_{\gamma\gamma}(E_{\gamma},z)+\tau_{\gamma N}(E_{\gamma})$, the differential photon flux at Earth, i.e., the number of photons with energy $E_\gamma$, per unit energy, per unit area, per unit time, can be calculated as
 \be
\frac{d\Phi_\gamma(E_\gamma)}{dE_\gamma}= \frac{1}{d^2}\int dV Q_\gamma(E_\gamma,z) \exp{\left[- \tau_{\gamma}(E_\gamma,z)\right] },\label{difflux} 
 \ee
{where $d$ is the distance from the source to Earth.}

\subsection{Neutrino emission}\label{subsec:neutrino}
Neutrinos arise from direct pion decays plus muon decays with a total emissivity  $$Q'_{\nu_\mu}(E',z)=
Q'_{\pi\rightarrow\nu_\mu}(E',z)+ Q'_{\mu\rightarrow\nu_\mu}(E',z),$$
{which correspond to the observer frame and transforms according to Eq. (\ref{Qgtrans}).}
The contribution from pion decays ($\pi^+\rightarrow \mu^+ \nu_\mu$, $\pi^-\rightarrow \mu^- \bar{\nu}_\mu$) is represented {in the co-moving frame} by
\begin{multline}
Q'_{\pi\rightarrow\nu_\mu}(E',z)= \int_{E'}^{\infty}dE_\pi
T^{-1}_{\pi,\rm d}(E_\pi)N'_\pi(E_\pi,z) \\ \times
\frac{\Theta(1-r_\pi-x)}{E_\pi(1-r_\pi)},
\end{multline}
with $x=E'/E_\pi$ and $T_{\pi,\rm d}=\gamma_\pi \, 2.6\times 10^{-8}{\rm s}$ .

The neutrino emissivity for muon decays ($\mu^- \rightarrow e^- \bar{\nu}_e \nu_\mu$,  $\mu^+ \rightarrow e^+ {\nu}_e \bar{\nu}_\mu$) can be calculated as
\begin{multline}
Q'_{\mu\rightarrow\nu_\mu}(E',z)= \sum_{i=1}^4\int_{E'}^{\infty}\frac{dE_\mu}{E_\mu} T^{-1}_{\mu,\rm
d}(E_\mu)N'_{\mu_i}(E_\mu,z) \\ \times \left[\frac{5}{3}-
3x^2+\frac{4}{3}x^3 
\right].
\end{multline}
In this expression, $x=E'/E_\mu$, $\mu_{1,2}=\mu^{-,+}_L$, $T_{\mu,\rm d}=\gamma_\mu \, 2.2\times 10^{-6}{\rm s}$, and
$\mu_{3,4}=\mu^{-,+}_R$ \citep[see][]{2007PhRvD..75l3005L}.
The differential neutrino flux can be computed using a expression analogous to Eq. (\ref{difflux}).

The synchrotron cooling of $\mu$s and $\pi$s, which is taken into account in our model, will be traduced on a lower neutrino emissivity as it was previously shown in \cite{2009A+A...493....1R}.

\subsection{{Secondary $e^+e^-$ production}}\label{subsec_pairs}
{We consider the production of secondary electrons and positrons in the jet through $p\gamma$, $pp$, and $\gamma\gamma$ interactions.
The inelastic $p\gamma$ collisions with can produce $e^-e^+$ pairs apart from pions as discussed above.
We consider this direct production of pairs through the injection function in the jet frame}
\begin{multline}
Q'_{e^\pm, p\gamma}(E',z)= 2 \int_{m_pc^2}^\infty d E_p' N_p'(E'_p,z) \\ \times {\omega'}_{p\gamma}^{(e)}(E'_p,z)\,\delta\left(E'-\frac{m_e}{m_p}E'_p\right),
\end{multline}
{where the $\omega_{p\gamma,e^\pm}(E'_p)$ is the collision frequency given by}
\begin{multline}
{\omega'}_{p\gamma}^{(e)}(E'_p,z)= \frac{c}{2\gamma_p^2}\int_{\frac{\epsilon_{\rm
th}^{(e)}}{2\gamma_p}}^{\infty} dE_{\rm ph}'\frac{n'_{\rm
ph}(E_{\rm ph}',z)}{{E_{\rm ph}'}^2}
\\ \times  \int_{\epsilon_{\rm
th}^{(e)}}^{2\epsilon\gamma_p} d \epsilon'
  \sigma_{p\gamma}^{(e)}(\epsilon')
  \; \epsilon'.
\end{multline}
{Here, the photopair cross section $\sigma_{p\gamma}^{(e)}$ is considered as in \citep{1990ApJ...362...38B}.}

As discussed above, the inelastic $pp$ and $p\gamma$ collisions produce pions that decay giving muons. The decay products of muons include electrons and positrons. The decay spectrum of the electrons and positrons are taken to be equal to that of muon neutrinos and anti-neutrinos \citep[see e.g.][]{1990cup..book.....G}. Thus, the injection of electrons and positrons from muon decay after $pp$ and $p\gamma$ interactions is considered following \cite{2002cra..book.....S}. 

As for the $e^+e^-$ pairs produced in $\gamma\gamma$ interactions, we consider the injection according to \cite{1983Afz....19..323A},
\begin{multline}
 Q'_{e^\pm, \gamma\gamma}(E',z)= \frac{3}{32}\frac{\sigma_{\rm T}c}{m_e c^2}\int_{\gamma_e}^\infty d\epsilon_\gamma \int_{\frac{\epsilon_\gamma}{4\gamma_e(\epsilon_\gamma-\gamma_e)}}^\infty d\omega \frac{n_\gamma(\epsilon_\gamma,z)}{\epsilon^3}
\\ \times \frac{n'_{\rm ph}(\omega m_ec^2,z)}{\omega^2}
\left\{ \frac{4\epsilon_\gamma^2}{\gamma_e(\epsilon_\gamma- \gamma_e)}\ln\left[\frac{4\gamma_e\omega(\epsilon_\gamma-\gamma_e)}{\epsilon_\gamma}\right] \right. \\
\left. -8\epsilon_\gamma\omega + \frac{4\epsilon_\gamma^3\omega-2 \epsilon_\gamma^2}{\gamma_e(\epsilon_\gamma-\gamma_e)}- \left(1-\frac{1}{\epsilon_\gamma\omega}\right)\frac{\epsilon_\gamma^4}{\gamma_e^2(\epsilon_\gamma-\gamma_e)^2}\right\},
\end{multline}
{where $\gamma_e={E'}/({m_e c^2})$, $\omega= {E'_{\rm ph}}/({m_e c^2})$ , and $\epsilon=E'_\gamma/(m_ec^2)$. The soft photon density $n'_{\rm ph}$ refers to the synchrotron radiation of electrons, while the density of higher energy photons from IC, $pp$, and $p\gamma$ interactions, is given by}
$$n'_\gamma(E'_\gamma,z)\approx 4\pi\frac{Q'_{\gamma, IC}(E'_\gamma,z)+ Q'_{\gamma, pp}(E'_\gamma,z)+  Q'_{\gamma, p\gamma}(E'_\gamma,z)}{  c/(z\tan{\xi_{\rm j}})+ t^{-1}_{\gamma\gamma}(E'_\gamma,z)}. $$
{Here, the annihilation rate of gamma-rays due to pair production is given by}
 \begin{multline}
 t_{\gamma\gamma}^{-1}(E'_\gamma,z)=2\pi \, c \int_{-1}^1dx'(1-x') \\ \times \int_{\frac{2m_e^2c^4}{E'_\gamma(1-x')}}^{\infty} dE'_{\rm ph} n'_{\rm ph}(E'_{\rm ph},z) \sigma_{\gamma\gamma}(E'_\gamma,E'_{\rm ph}).
 \end{multline}

The distribution of secondary electrons and positrons in the jet is found solving the transport equation. This allows us to calculate the synchrotron and IC emission produced by these particles.

\section{Application to Centaurus A}\label{sec:AplCenA}

\begin{figure*}[!htp]
\includegraphics[trim = 6mm 5mm 4mm 45mm, clip,width=\linewidth,angle=0]{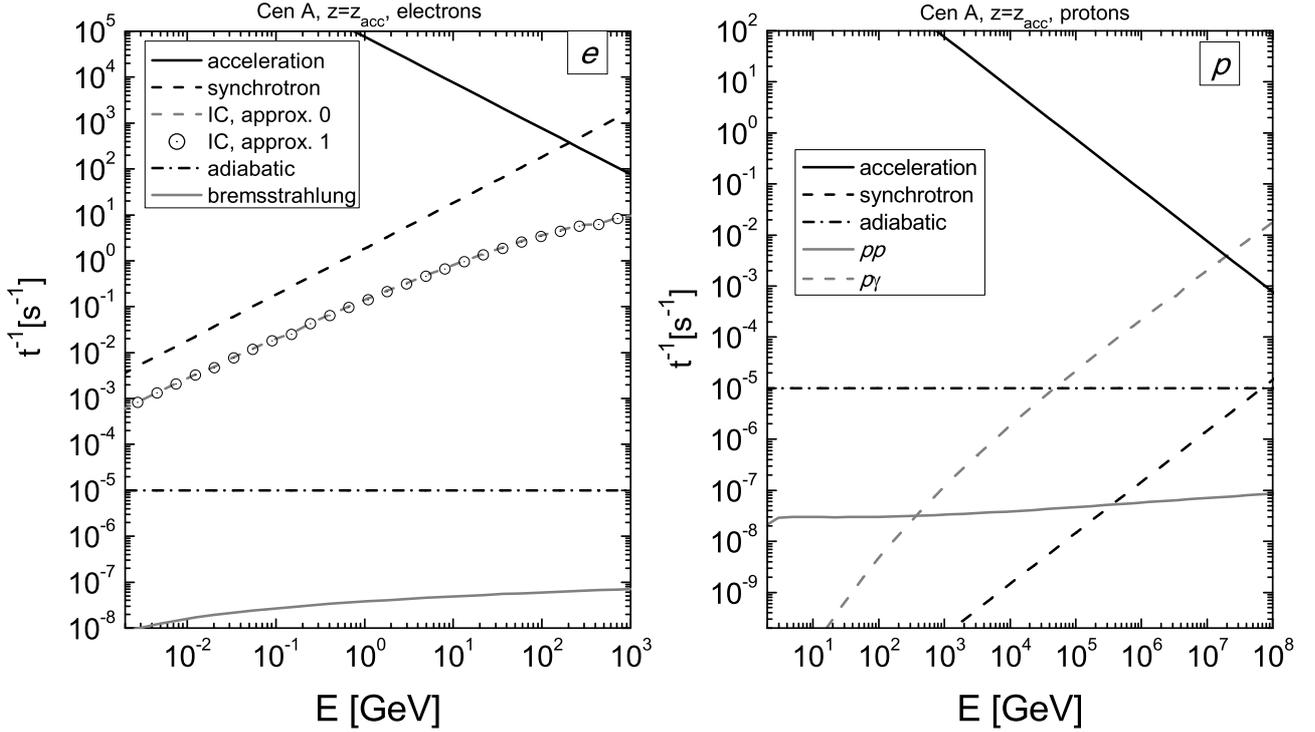}
\caption{Accelerating and cooling rates for electrons ({\it left})
and for protons ({\it right}) {at a distance $z = z_{\rm
acc}$ from the central engine of Cen A. In the case of electrons, the light-grey dashed line correspond to the first calculation of the IC cooling rate taken into account only synchrotron cooling to obtain the electron distribution. A second approximation to the electron distribution leads to the IC cooling rate indicated in white circles.}} \label{tep_cena}
\end{figure*}

\begin{figure*}[!htp]
\includegraphics[trim = 0mm 0mm 0mm 0mm, clip,width=\linewidth,angle=0]{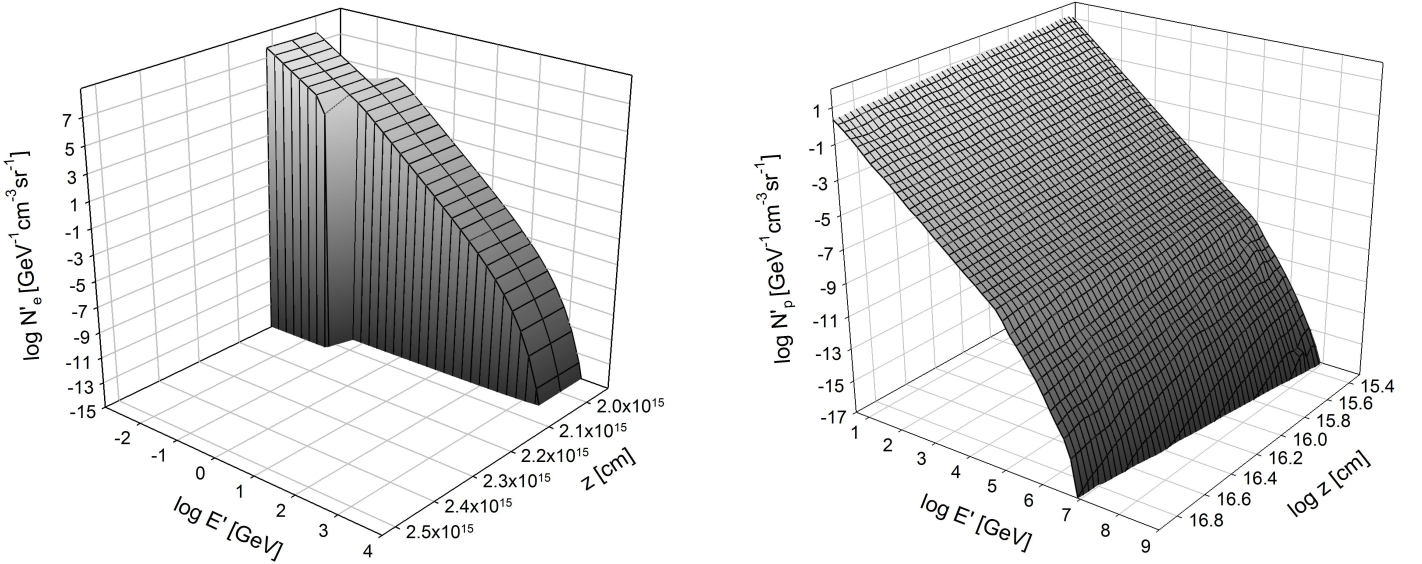}
\caption{Distributions of primary particles as function of the energy and the distance to the core inside the Cen\,A jet.{\it Left}: electrons;~{\it right}: protons.}
\label{NeNp_cena}
\end{figure*}

In this section we present the results for Centaurus\,A, which were obtained applying our model with the set of parameters listed on Table \ref{Tab:paramsCena}.
The cooling rates for high energy electrons and protons for this configuration are shown in Fig.\,\ref{tep_cena}, while the obtained electron and proton distributions $N'_{e}(E',z)$ and $N'_p(E',z)$ are presented in Fig.\,\ref{NeNp_cena}.

   \begin{table}[!htp]
      \caption[]{Model parameters for Cen A}
         \label{Tab:paramsCena}
     $$
         \begin{array}{p{0.68\linewidth}l}
            \hline
            \noalign{\smallskip}
            Parameter      &  {\rm Value}  \\
            \noalign{\smallskip}
            \hline
            \noalign{\smallskip}
            $M_{\rm bh}$: black hole mass & 10^{8} M_{\odot}\\
            $R_{\rm g}$: gravitational radius & 1.47 \times 10^{13} {\rm cm} \\
            $L_{\rm j}^{\rm(kin)}$: jet kinetic power at $z_0$ & 6.28\times 10^{44} {\rm erg \ s^{-1}}   \\      
            $q_{\rm j}$: ratio $2 L_{\rm j}^{\rm (kin)}/L_{\rm Edd}$ & 0.1 \\
            $\Gamma_{\rm b}(z_0)$:  bulk Lorentz factor of the jet at $z_0$  & 3   \\
            $\theta$: viewing angle &  30^\circ     \\
            $\xi_{\rm j}$: jet's half-opening angle &  2.5^\circ \\
            $q_{\rm rel}$: jet's content of relativistic particles & 0.05 \\
            $a$: hadron-to-lepton power ratio     & 0.025        \\
            $z_0$: jet's launching point   & 50 \ R_{\rm g} \\
            $q_{\rm m}$: magnetic to kinetic energy ratio at $z_{\rm acc}$ & 0.38\\
            $z_{\rm acc}$: injection point & 132 \, R_{\rm g}  \\
            $\Delta z$: size of injection zone & 2\,z_{\rm acc} \tan\xi_{\rm j}=11.5 \ R_{\rm g}  \\
            $B(z_{\rm acc})$: magnetic field at $z_{\rm acc}$ & 3065 \, {\rm G} \\
            $m$: index for magnetic field dependence on $z$ & 1.5\\
            $s$: injection spectral index & 1.8\\
            $\eta$: acceleration efficiency & 10^{-2}\\
            $E_p^{\rm (min)}$: minimum proton energy & 3 \,{\rm GeV} \\
            $E_e^{\rm (min)}$: minimum electron energy & 0.1 \,{\rm GeV} \\
            $E_p^{\rm (max)}$: maximum primary proton energy & 2 \times 10^7\,{\rm GeV} \\
            $E_e^{\rm (max)}$: maximum primary electron energy & 200 \,{\rm GeV} \\
            $N_H$: column dust density & 10^{23} \,{\rm cm^{-2}}\\
            $n_c({z_{acc})}$: cold matter density inside the jet at $z_{\rm acc}$ & 3 \times 10^{8} \,{\rm cm^{-3}} \\
            \noalign{\smallskip}
            \hline
         \end{array}
     $$
   \end{table}

We can see in Fig.~\ref{tep_cena} that the relevant energy looses are due to synchrotron cooling for electrons and $p\gamma$ interactions for protons. Since the plot corresponds to the injection zone ($z=z_{\rm acc}$), $p\gamma$ interactions are favored by a large density of target photons corresponding to the synchrotron emission of electrons. These photohadronic interactions are not so important for protons outside the injection zone.

It is important to notice that the maximum energy achievable for the protons in this context is of $2\times 10^{7}$ GeV. These protons can not account for the UHE cosmic rays detected by the Pierre Auger Observatory in the direction of Cen\,A but it gives a hint to other explanations. If neutrons of the same energy are produced by pion photoproduction inside the jet, these could been beamed along the jet and decay into protons near the outer radio lobes, where they would be re-accelerated up to the observed energies by Auger. Other possibilities were proposed by \cite{2009A+A...506L..41R}, with particles that go through shear acceleration inside the jet, and by \cite{1996APh.....5..279R}, who considered that the UHE cosmic rays are produced at the extended radio lobes. If this is the case, it has been estimated by \cite{2009MNRAS.393.1041H} that the minimum jet power required is roughly $L_{\rm j}^{\rm(kin)}>3.5\times 10^{43}{\rm erg \, s}^{-1}$, which is consistent with the value assumed here.

Fig.~\ref{NeNp_cena} shows that the electron distribution drops quickly with $z$ at the end of the injection zone,  while the proton distribution is also important further along the jet. This is a consequence of the different energy loss rates of the particles: electrons cool very rapidly emitting synchrotron radiation, and protons lose energy at a much lower rate as they propagate along the jet.

\begin{figure*}[!htp]
\includegraphics[trim = 0mm 20mm 5mm 0mm, clip,width=17cm,angle=0]{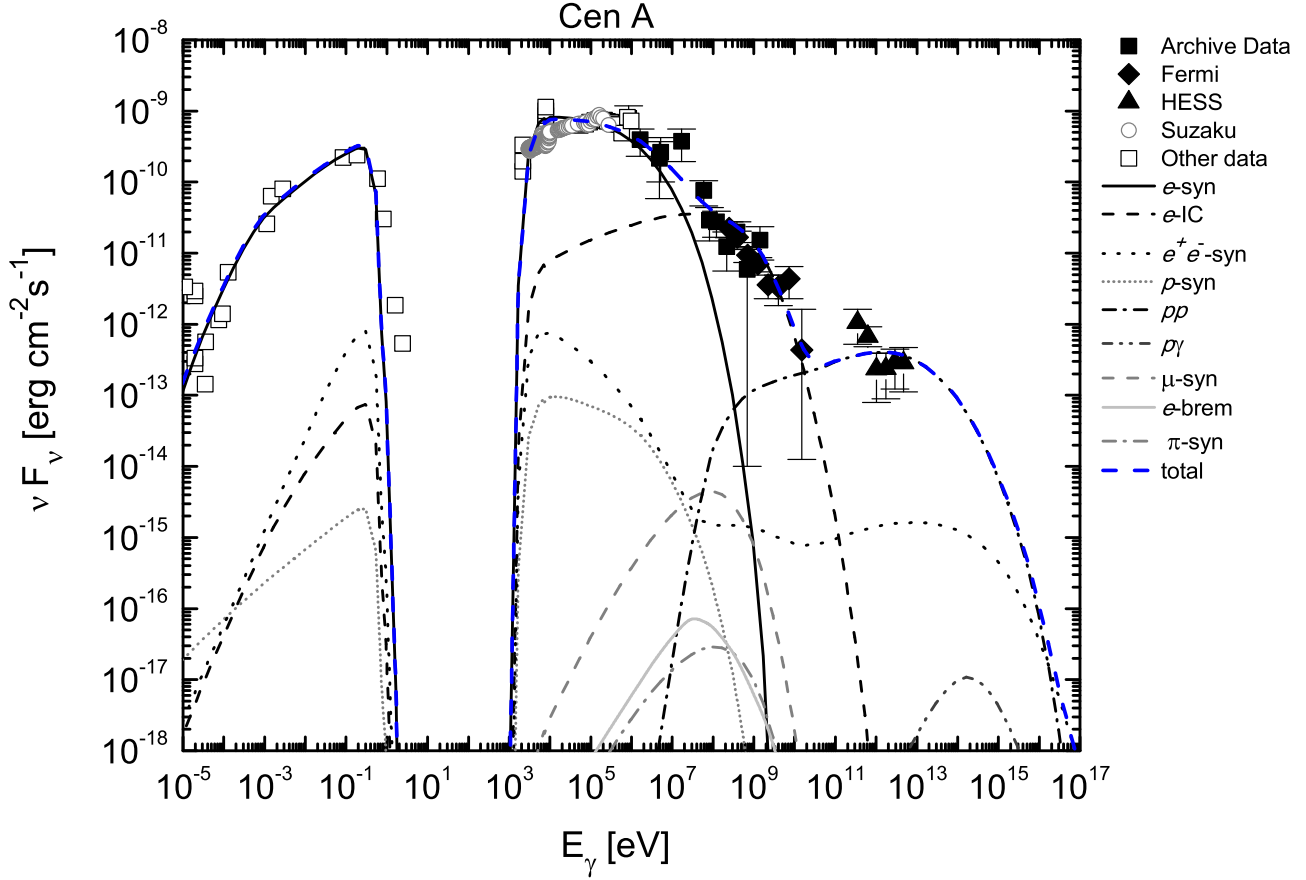}
\caption{Model output for the SED of Cen\,A. The different emission processes are indicated, together with the total output. The recent observational data is also included: {\it Fermi}/LAT (black filled diamons) \citep{2010ApJ...719.1433A} and HESS spectra (black filled triangles). The rest of the data points correspond to the references given in the text.}
\label{SED_cena}
\end{figure*}

The SED of Cen\,A (see Fig.~\ref{SED_cena}) includes the HESS spectrum in the VHE range together with data from {\it CGRO}/COMPTEL \citep{1998A+A...330...97S}, {\it RXTE} and {\it INTEGRAL} \citep{2006ApJ...641..801R}. We also include data from {\it HST}/NICMOS and WFPC2 \citep{2000ApJ...528..276M}, SCUBA at 800 $\mu$m  from \citep{1993MNRAS.260..844H}, ISO and SCUBA (450 $\mu$m and 850 $\mu$m) \citep{1999A+A...341..667M}, {\it XMM}-Newton, {\it Chandra} spectra and the {\it Suzaku} data presented in \citep{2010ApJ...719.1433A}. The {\it Fermi}/LAT correspond to Cen\,A core as given in \citep{2010ApJ...719.1433A}, and, finally, we also show the old data from the NASA Extragalactic Database.

The {\it Fermi}/LAT analysis revealed that the high-energy spectrum is non-variable over the first ten months of scientific operation of the instrument. This steady behavior is also supported by the HESS experiment, which also reported a constant flux from Cen\,A even though {\it CGRO} data shows some variability with at least two emission states during the period 1991-2000.

Since the set of measurements composing the SED of Cen\,A is quite inhomogeneous in time and angular resolution, so one should be very careful in attempting to interpret any fit of all the spectrum simultaneously. In particular, the data that define the bump in the hard-X-rays ($\sim$0.1 MeV) have been taken more than 10 years ago with a poor angular resolution and long integration times. The lack of a good spatial resolution makes impossible to distinguish the emission components (jet, nucleus or other radiation sources). Furthermore, we note a discrepancy between the flux normalization of {\it Fermi}/LAT and HESS, which is not yet fully understood.

Nevertheless, we have obtained using our model a spectral energy distribution which is basically consistent with the multi wavelength emission from Cen\,A.

An important role is played by absorption due to photoionization interactions in the surrounding dust. {This arises due to the large value adopted for the column density of neutral hydrogen, $N_H=10^{23}{\rm cm}^{-2}$} \citep[see, e.g.][]{2004ApJ...612..786E,2008A+A...485L...5M}. A drastic modulation is then imprinted in the electron synchrotron spectrum, which is responsible for the whole emission in the broadband range $10^{-5}-10^{7}$eV. Such situation is possible if electrons can be efficiently accelerated ($\eta =0.01$) to high energies with a rather flat spectral index ($s=1.8$).
The internal $\gamma\gamma$ absorption does not modify significantly the $pp$ contribution of gamma-rays, since it is only important within the injection zone, which is a negligible region of the jet compared to the one in which $pp$ collisions occur (see Fig. \ref{NeNp_cena}).

\begin{figure} [htp]
\includegraphics[trim = 50mm 05mm 70mm 60mm, clip,width=\linewidth,angle=0]{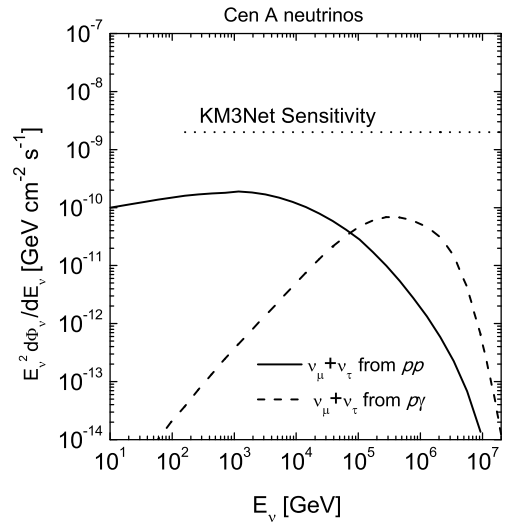}
\label{FigE2nuCena}
\caption{Differential neutrino flux weighted by the squared energy as predicted by the model for Cen A. {\it Dotted line}: Approximate KM3Net sensitivity for 1 year operation.}
\end{figure}

We evaluate the accompanying neutrino output using the same set of parameters. The obtained differential flux, weighted by the squared energy is plotted in Fig.~\ref{FigE2nuCena} together with the estimated sensitivity of KM3Net for one year of operation \citep{2009NIMPA.602...40F}.
We can see in this figure that the neutrino signal produced by $pp$ and $p\gamma$ interactions would not be observable by KM3NeT detector in just one year of observation. It is to be noticed, however, that the sensitivity level shown in the plot actually corresponds to a neutrino spectrum with an $E_\nu^{-2}$ dependence, which is a bit steeper than our case. Hence, we can expect that the actual sensitivity for our flux will be better than what is shown here, so with two or three years of data taking it would be possible to achieve detection.

\section{Application to M87}\label{sec:AplM87}
\begin{figure*}
\includegraphics[trim = 5mm 05mm 10mm 45mm, clip,width=\linewidth,angle=0]{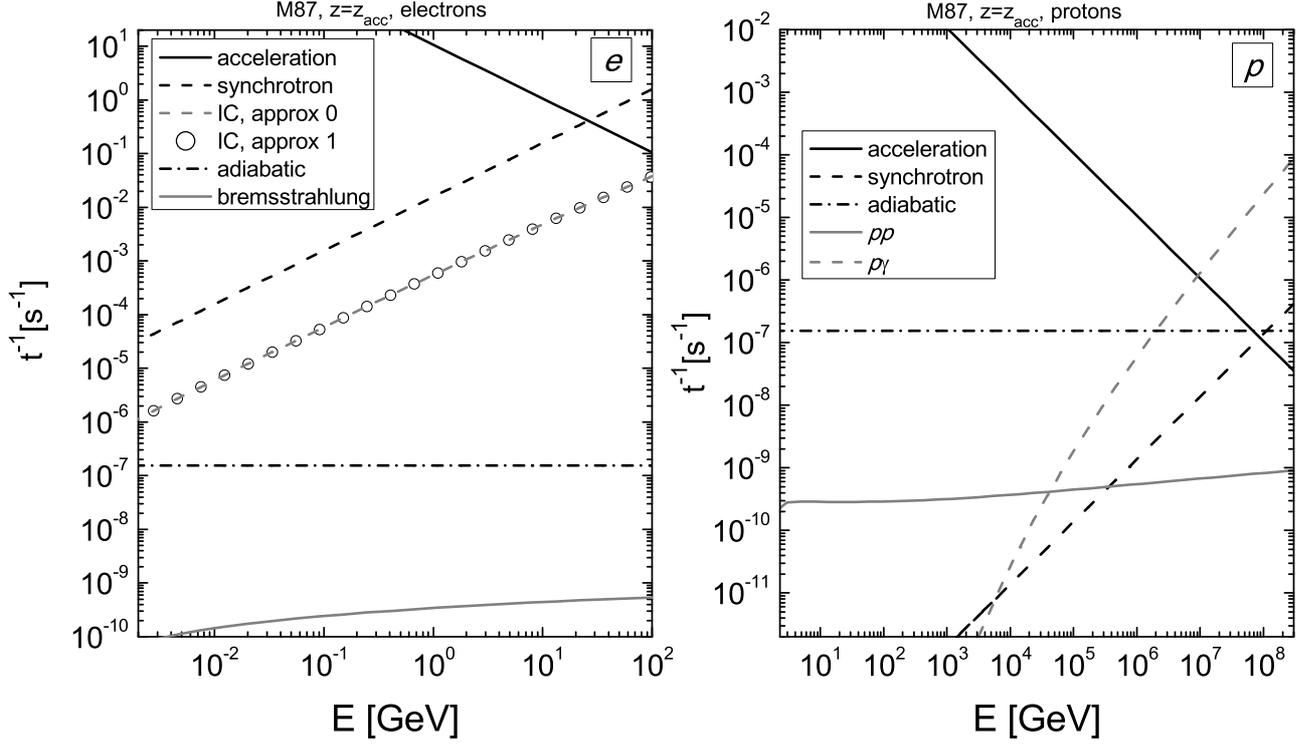}
\caption{Accelerating and cooling rates for electrons ({\it left}) and for protrons ({\it right}) {at a distance $z = z_{\rm acc}$ from the central engine of M87.}}
\label{tep_m87}
\end{figure*}

\begin{figure*}[!htp]
\includegraphics[trim = 0mm 0mm 0mm 0mm, clip,width=\linewidth,angle=0]{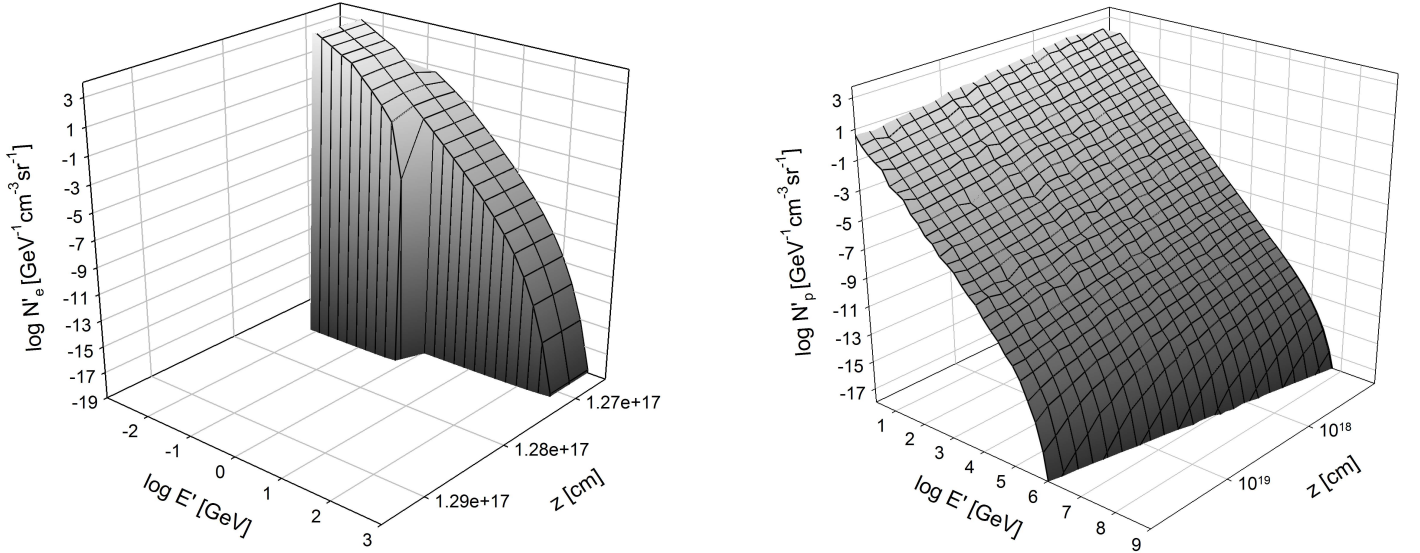}
\caption{Distributions of primary particles as function of the energy and the distance to the core inside the M87 jet.{\it Left}: electrons;~{\it right}: protons.}
\label{NeNp_m87}
\end{figure*}

In the case of M87, we have used the set of parameters listed on Table \ref{Tab:paramsM87} as well as the multi wavelength data available. We show the cooling rates for high energy electrons and protons in Fig.~\ref{tep_m87}, and the electron and proton distributions $N'_{e}(E',z)$ and $N'_p(E',z)$ in Fig.~ \ref{NeNp_m87}.
   \begin{table}[htp]
      \caption[]{Model parameters for M87.}
         \label{Tab:paramsM87}
     $$
         \begin{array}{p{0.68\linewidth}l}
            \hline
            \noalign{\smallskip}
            Parameter      &  {\rm Value}  \\
            \noalign{\smallskip}
            \hline
            \noalign{\smallskip}
            $M_{\rm bh}$: black hole mass & 6 \times 10^{9} M_{\odot}\\
            $R_{\rm g}$: gravitational radius & 8.87 \times 10^{14} {\rm cm} \\
            $L_{\rm j}^{\rm (kin)}$: jet power at $z_0$ & 1.9\times 10^{46} {\rm erg \ s^{-1}}   \\
            $q_{\rm j}$: ratio $2 L_{\rm j}^{\rm (kin)}/L_{\rm Edd}$ & 0.05 \\
            $\Gamma_{\rm b}(z_0)$: bulk Lorentz factor of the jet at $z_0$  & 4   \\
            $\theta$: viewing angle &  30^\circ     \\
            $\xi_{\rm j}$: jet's half-opening angle &  1.5^\circ \\
            $q_{\rm rel}$: jet's content of relativistic particles & 0.05 \\
            $a$: hadron-to-lepton power ratio     & 40         \\
            $q_{\rm m}$: magnetic to kinetic energy ratio at $z_{\rm acc}$ & 0.35\\
            $z_0$: jet's launching point   & 50 \;R_{\rm g} \\
            $z_{\rm acc}$: injection point & 143 \; R_{\rm g}  \\
            $\Delta z$: size of injection zone & z_{\rm acc}\tan\xi_{\rm j}/4 = 11.5 \, R_{\rm g}  \\
           $B(z_{\rm acc})$: magnetic field at $z_{\rm acc}$ & 407 \, {\rm G}\\
            $m$: index for magnetic field dependence on $z$ & 1.5\\
            $s$: injection spectral index  & 2.4\\
            $\eta$: acceleration efficiency &  1.5\times 10^{-5}\\
            $E_p^{\rm (min)}$: minimum proton energy & 3 \,{\rm GeV} \\
            $E_e^{\rm (min)}$: minimum electron energy& 65  \,{\rm MeV} \\
            $E_p^{\rm (max)}$: maximum primary proton energy & 9\times 10^6 \, {\rm GeV} \\
            $E_e^{\rm (max)}$: maximum primary electron energy & 25 \, {\rm GeV} \\
            $N_H$: column dust density & 2 \times 10^{20}{\rm cm^{-2}}\\
            $n_c({z_{\rm acc}})$: cold matter density inside at $z_{\rm acc}$ & 3.9 \times 10^{6} \,{\rm cm^{-3}} \\

            \noalign{\smallskip}
            \hline
         \end{array}
     $$
   \end{table}

We can see in Fig.~\ref{tep_m87} that the relevant energy looses are due to synchrotron cooling of electrons and protons. Protons also cool significantly through adiabatic losses, and although $pp$ interactions are not dominant, they can give an important radiative output. From the balance between this radiative loss rate and the acceleration rate, the maximum energy achievable for the protons inside the jet is about $9\times 10^{6}$ GeV. These protons do not escape from the source. The only possibility to obtain a detectable cosmic rays flux at the earth coming from M87 should depend on neutrons produced by pion photoproduction processes and relativistically beamed along the jet direction and Doppler boosted in energy.

As in the case of Cen\,A, Fig.~\ref{NeNp_m87} shows that electrons radiate only at the base of the jet loosing all their energy rapidly while protons survive longer inside the jet, emmiting enough radiation to account for the VHE $\gamma$ observed emission.

\begin{figure*}[!htp]
\includegraphics[trim = 0mm 15mm 4mm 0mm, clip,width=17cm,angle=0]{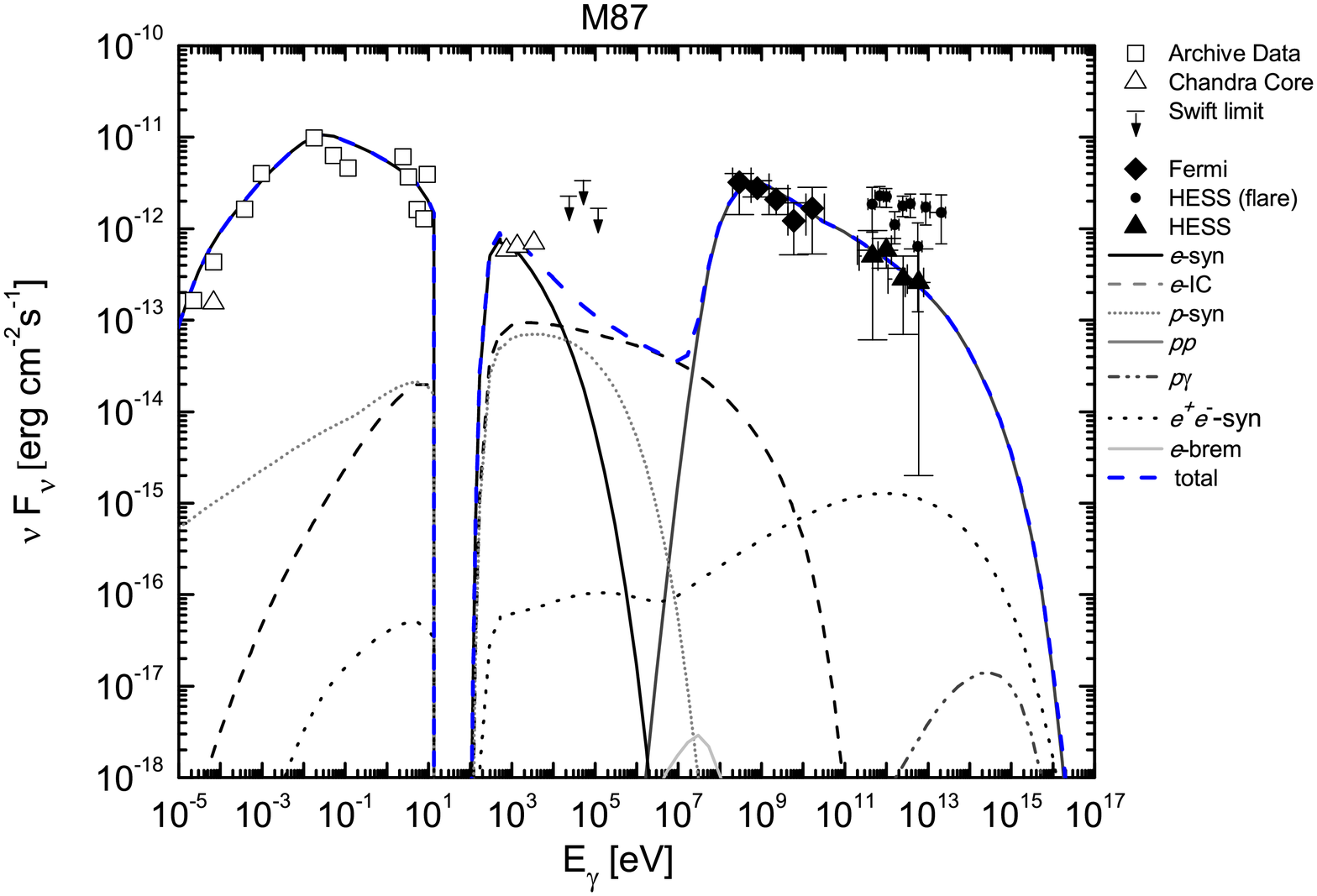}
\caption{M87 SED taking into account the recent {\it Fermi}/LAT, HESS spectra in two different states and historical multi wavelength data recompiled in \citep{2009ApJ...707...55A}. Model output for different processes and particles and the total output are also shown. {The model fits the quiescent state.}}
\label{sed-M87}
\end{figure*}

In Fig~\ref{sed-M87} we plot the contributions to the SED of M87 obtained with the different emitting proceeses in our model. We also show the corresponding multi wavelength observational data.  The HESS spectrum included in the VHE range corresponds to a low state from the 2004 observing season \citep{2006Sci...314.1424A}. Data from Fermi/LAT  \citep{2009ApJ...707...55A}, 2009 MOJAVE VLBA 15 GHz \citep{2004ApJ...609..539K} and Chandra X-ray measurements of the core  \citep{2009ApJ...699..305H} are also shown. We also include the 3$\sigma$ upper limits of the integrated emission in three hard X-ray bands  based on the Swift/BAT dataset \citep{2009ApJ...699..603A}. Historical measurements of the core are also plotted: VLA 1.5, 5, 15 GHz \citep{1991AJ....101.1632B}, IRAM 89 GHz \citep{1996A+A...309..375D}, SMA 230 GHz \citep{2008ApJ...689..775T}, Spitzer 70, 24 μm \citep{2007ApJ...655..781S}, Gemini 10.8 μm \citep{2001ApJ...561L..51P} and HST optical/UV \citep{1996ApJ...473..254S}.

We have obtained, using our model, a spectral energy distribution which is basically consistent with the multi wavelength emission from M87. The spectrum detected by the Fermi/LAT collaboration connects smoothly with the low-state TeV spectrum detected by HESS. Our model reproduces this connection effectively with the $\gamma$-ray emission produced by $pp$ collisions.
Contrary to the case of Cen\,A, this radiative output for M87 is possible if particles are not so efficiently accelerated ($\eta =1.5\times 10^{-5}$) to high energies with a rather steep spectral index ($s=2.4$).

The flaring state, which is not shown in Fig. \ref{sed-M87},  can still be reproduced in our model by $\gamma$-rays from $pp$ interactions if we assume a flatter spectrum of injected protons and a negligible power in relativistic electrons.

The accompanying high energy neutrino flux was calculated using also the set of parameters of Table \ref{Tab:paramsM87}. We show the differential neutrino flux weighted by the squared energy as a function of energy in Fig. ~\ref{FigE2nuM87}, where we also plot the estimated sensitivity of IceCube for one year of operation \citep{2004APh....20..507A}.

\begin{figure}[!htp]
\includegraphics[trim = 50mm 15mm 70mm 45mm, clip,width=\linewidth,angle=0]{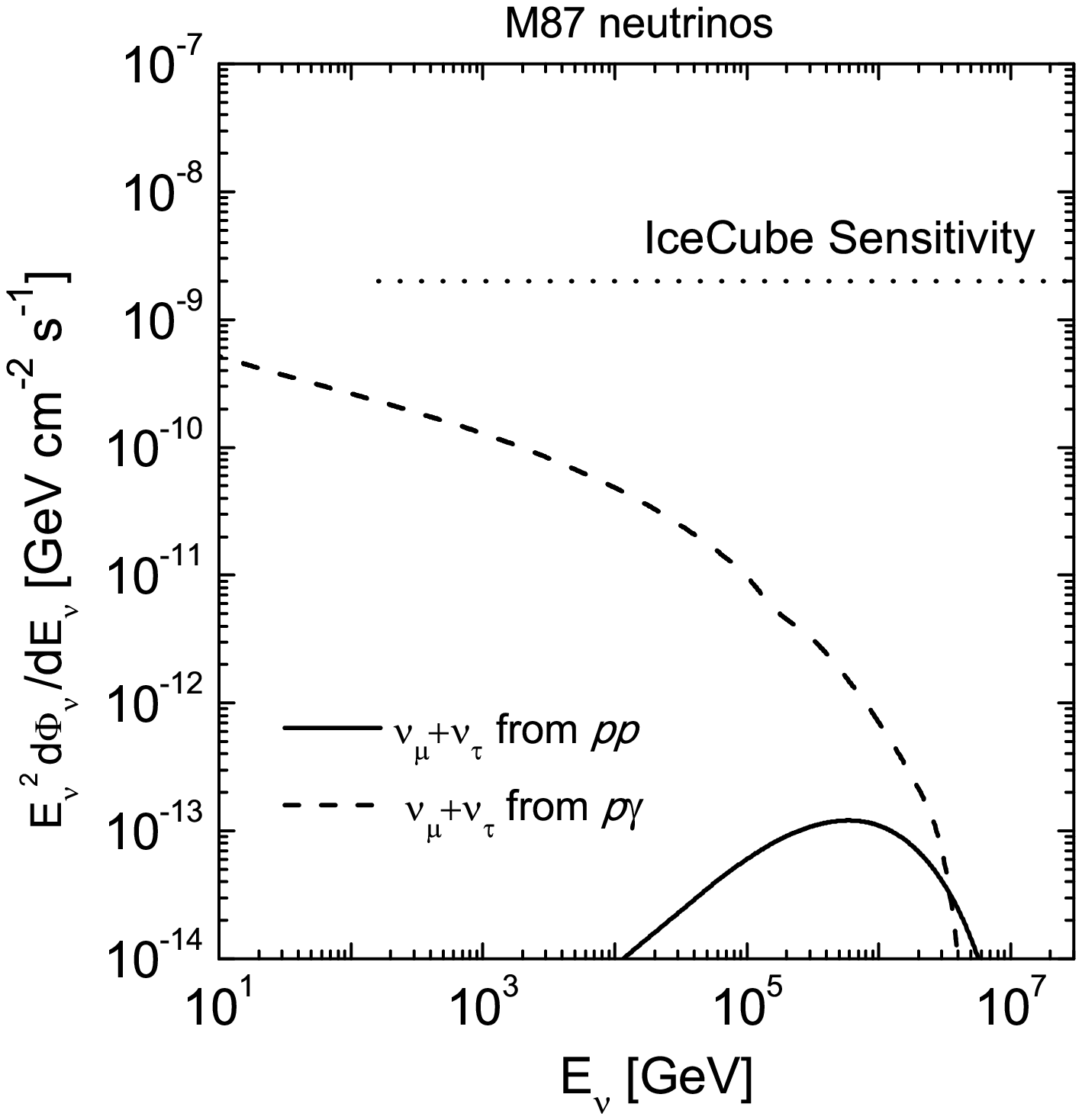}\label{FigE2nuM87}
\caption{Differential neutrino flux weighted by the squared energy as predicted by the model for M87. {\it Dotted line}: Approximate IceCube sensitivity for 1 year operation.}
\label{FigE2nuM87}
\end{figure}

As can be seen been from Fig. \ref{FigE2nuM87}, the neutrino signal associated to M87 would be hard to detect. However, the sensitivity shown corresponds to a neutrino spectrum with an $E_\nu^{-2}$ dependence, i.e., not so steep as our predicted one. Hence, a more detailed study could help to assess the real detectability of this type of signal \citep{2009PhRvD..79d3013N, 2009PhRvD..80h3008N}.

\section{Conclusions}\label{sec:concl}

We have presented here a lepto-hadronic model describing the particle propagation and interactions in the inclined jets of radiogalaxies. The scenario includes an acceleration zone placed near the jet base, where relativistic particles are injected. The location of this injection zone is fixed in the jet at a distance $z_{\rm acc}$ from the black hole by requiring that the magnetic energy density is less prominent than the bulk kinetic density. This favors shock formation.

One very important aspect of the present model is that it does not consist of a one-zone treatment, since particles are allowed to convect along the jet, away from the injection zone. To obtain the particle distributions, we have used a stationary one-dimensional transport equation that accounts for energy losses, escape, and convection of particles. The resulting distribution of primary electrons is important mainly in the injection zone, since they undergo a rapid synchrotron cooling. Instead, protons are convected away along the jet, loosing energy gradually (mainly by adiabatic cooling, synchrotron radiation, and $pp$ interactions). Also they undergo significant $p\gamma$ collisions with the electron-synchrotron photons at the injection zone. The secondary pions and muons produced in the hadronic interactions are also considered, and their distributions are found using the mentioned type of transport equation. This allows to account for the synchrotron cooling suffered before the decay of this secondary, transient particles.

We have found that the radiative output obtained can account for much of the observational data for the two radiogalaxies M87 and Cen\,A. As seen in Figs.~\ref{SED_cena} and ~\ref{sed-M87}, the model allows for the possibility of explaining the VHE emission by means of $pp$ interactions. The hard X-ray data corresponds to electron synchrotron radiation whereas the soft-$\gamma$ emission is well reproduced by proton synchrotron and IC emission. {Previous hadronic models (e.g. the synchrotron proton blazar of \cite{2003APh....18..593M}) could also fit the Fermi energy range for M87 and the hard spectrum measured by HESS (consistent with VERITAS and MAGIC measurements), according to what it is mentioned in \cite{2009ApJ...707...55A}}. A previous one-zone lepto-hadronic model for Cen\,A was proposed by \cite{2009AIPC.1123..242O} just before HESS and Fermi/LAT data was available.

In the present work, we have included the effects of absorption due to photoionization interactions in the surrounding {medium}. This causes a drastic modulation in the electron synchrotron spectrum, especially for Cen\,A, which is responsible for the whole emission in the broadband range $\sim 10^{-5}-10^{7}$eV. In the case of M87, this type of absorption has a minor impact, since the column density is much less ($N_H=2 \times 10^{20}{\rm cm}^{-2}$, \cite{1996ApJ...458L...5L}).
The internal absorption of gamma-rays is also taken into account, being important only within the injection zone, where the synchrotron emission of primary electrons provide an important absorbing target.

It is important to note that the same model can be used to describe different types of jet. In terms of the broadband photon emission, M87 data shows a similar luminosity level for high and the low energy ranges. Since the VHE emission is basically determined by $pp$ interactions, we need a large power to be injected in relativistic protons as compared to that injected in electrons. This is because electrons cool completely, mainly by synchrotron emission, and protons undergo an important adiabatic cooling. Hence, in the case of M87, we need a high value for the proton-to-electron power ratio: $a=40$. A different situation arises for Cen A, where the highest luminosities correspond to energies $E_{\gamma}<10^7{\rm eV}$, and the HE and VHE luminosities are comparatively lower.
In this case, we need {more power to be injected in electrons than in protons}  ($a=2.5\times 10^{-2}$) to allow the broadband electron-synchrotron spectrum to approximately account for all the data for $E_{\gamma}<10^7{\rm eV}$. Then, for Cen\,A a lower luminosity in protons is enough to reach the level of the observed HE and VHE radiation through proton-synchrotron and $pp$ interactions, respectively.
A further difference between Cen\,A and M87 is in the slope of the injected particle distributions: for Cen\,A, we have a quite flat injection ($s=1.8$), while for M87 we find a steeper injection ($s=2.4$). Also the acceleration efficiency is greater for Cen A than for M87. This also has an impact in the neutrino spectrum produced, which for $E_\nu> 1 $TeV appears more difficult to be detected for M87 than for Cen\,A.

Finally, we remark that a possible improvement of the present treatment, which is left for future work, should be the inclusion of a time-dependent injection to account for flaring states.

\begin{acknowledgements}
This work had the support from the GdR-PCHE, the Observatory of Paris, and CONICET (PIP 112-200801-00587, and  PIP 112-200901-00078). The authors would like to thank Andreas Zech, Jean-Philippe Leanain, Catherine Boisson and Helene Sol for the very fruitful discussions and the interesting contributions to this work, especially for Cen\,A.  We also thank Gabriela Vila and Chlo\'e Guennou for useful discussions on M87 and jet physics.
\end{acknowledgements}

\appendix

\section{Bulk Lorentz factor evolution}\label{appdx_gam}
\begin{figure}[!htp]
\includegraphics[trim = 0mm 5mm 0mm 30mm, clip,width=\linewidth,angle=0]{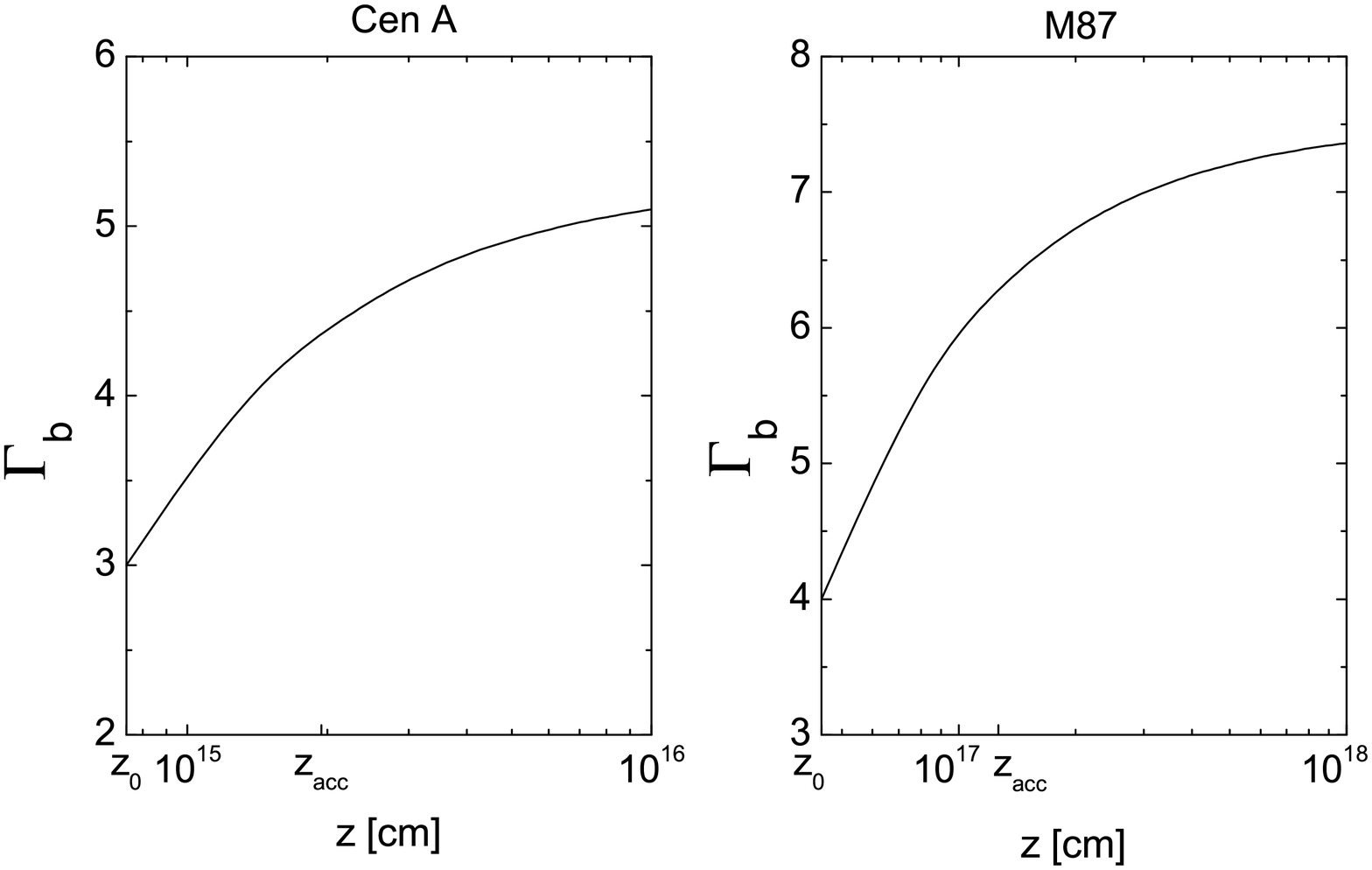}
\caption{Bulk lorentz factor of the jet as a function of the distance to the black hole obtained for Cen A (left panel) and M87 (right panel).}\label{FigGammab}
\end{figure}

{The sum of the kinetic plus magnetic energy in the jet is}
 \be
  dz \pi z^2\chi^2\left[ \rho_{\rm k}(z) + \rho_{\rm m}(z) \right]= 2 \rho_{\rm m}(z_0)  dz \pi z_0^2 \chi^2
 \ee
{where} $\chi=\tan\xi_{\rm j}$. {Then, }\be \rho_{\rm k}(z)=
\frac{B_0^2}{8\pi} \left(\frac{z_0}{z}\right)^2\left[2 -
\left(\frac{z_0}{z}\right)^{2m-2}\right]. \ee {Since} \be \rho_{\rm
k}(z)= \left[\Gamma_{\rm b}(z)- 1\right]\frac{\dot{m}_{\rm
j}c^2}{v_{\rm b}(z) z^2\tan^2\xi_{\rm j} } \ee {and given that} \be
v_{\rm b}=c\sqrt{1- \frac{1}{\Gamma_{\rm b}^2(z)} }, \ee {we obtain}
\be \frac{\Gamma_{\rm b}(\Gamma_{\rm b}-1)}{\sqrt{\Gamma_{\rm
b}^2-1}}= \frac{B_0^2 \chi^2 z_0^2}{8\dot{m}_{\rm
j}c}\left[2-\left(\frac{z_0}{z}\right)^{2m-2}\right]\equiv\sqrt{A_z}.
\ee {Squaring, we obtain the equation} \be \frac{\Gamma_{\rm
b}^2(\Gamma_{\rm b}-1)}{{\Gamma_{\rm b}+1}}= A_z \ee {which has the
analytic solution}
 \begin{multline}
\Gamma_{\rm b}(z)=\frac{1}{3} \left[18 A_z+3 \sqrt{3}
   \sqrt{11 A_z^2+A_z-A_z^3}+1\right]^{1/3}+\frac{1}{3} \\
+\frac{3 A_z+1}{3 \left[18 A_z+3 \sqrt{3} \sqrt{11
   A_z^2+A_z-A_z^3}+1\right]^{1/3}}.
 \end{multline}
For Cen A and M87 we obtain the results shown in Fig. \ref{FigGammab}.

\section{Absorption by photoionization in the surrounding medium }\label{appdx_rec}
 {Here we discuss on the absorption of photons once they espape from the jet and propagate in the ambient medium filled with hydrogen gas. In principle, photons with energies above the Lyman edge, $13.6$ eV, can ionize the neutral hydrogen atoms, and the recombination of electrons and protons can also occur. It is expected that the vicinity of the jet will be highly ionized, but since the emitted radiation can not ionize an infinite volume of gas, then the gas must become neutral at some point \citep[see e.g.][]{1978ppim.book.....S}.}

{A crude estimation of the size of the ionized zone can be attempted assuming a spherical symmetry of the emitting zone and of the radiation emitted. Considering that the gas extends upto a distance $R_H$ from the emitting zone, we adopt an uniform density of hydrogen $n_H=N_H/R_H$. This is actually the density of ionized hydrogen plus that of neutral hydrogen:}
\be
 n_H=n_p+ n_{HI}.
\ee
{The size of the ionization zone can then be estimated through the so-called Str\"{o}mgren radius $r_{\rm S}$ \citep[e.g.][]{1939ApJ....89..526S}, which is defined by equating the rate of ionizing photons $\dot{N}_\gamma$ to the number of recombinations per unit time:}
 \be
\dot{N}_\gamma= \alpha_{\rm rec}^{(2)} n_e^2  \frac{4\pi}{3}r_{\rm
S}^3.
 \ee
{Here, the number density of free electrons is taken to be $n_e\approx n_H$ within the ionized zone, and the rate of ionizing photons is
$$ \dot{N}_\gamma\approx \int_V dV \int_{E_\gamma>13.6 \rm{eV} } dE 4\pi Q_{\gamma, {\rm e-syn}},$$}
{which in our cases corresponds mainly to synchrotron radiation of electrons. The recombination coeficient can be taken as $\alpha_{\rm rec}^{(2)}\approx 2.6\times 10^{-3}{\rm cm}^{3}{\rm s}^{-1}$ for $T=10^4 {\rm K}$ and neglecting recombinations to the energy level $n=1$, since this state is ionized again {\it on the spot} \citep[e.g.][]{1999agnc.book.....K,2009A+A...507.1327H}). It can be checked that in the cases studied here, the Str\"{o}mgren radius is less than $10$ pc, which is assumed to be much less than $R_H$ . If the ionization fraction is $x=n_p/n_H$, then the density of neutral hydrogen is $n_{H}(1-x)$, which is an increasing function of the distance within the ionized region. The optical depth at a distance $R$ from the source and along the photon path can be integrated as}
\be
\tau_{\gamma N}(E_\gamma, R)= \int_{0}^{R}dl_\gamma \, n_{H}(1-x)\sigma_{\gamma N}(E_\gamma)
\ee
{According to \cite{1972ApJ...177L..69P}, the fraction of neutral gas reaches unity at the Str\"{o}mgren radius or even less in the prensence of dust. If the $r_S\gg R_H$ we can approximate the optical depth at $R_H$ as}
\be
 \tau_{\gamma N}(E_\gamma)\approx \sigma_{\gamma N}(E_\gamma)N_H,
\ee
which is the expression we use in this work.

\bibliographystyle{aa}  
\bibliography{14998}

\end{document}